\documentclass[11pt, twoside]{article}
\usepackage{soul}
\usepackage{xcolor}
\usepackage[most]{tcolorbox}
\tcbset{highlightstyle/.style={colback=yellow!30, colframe=yellow!70!black, boxrule=0mm, arc=0mm, top=2mm, bottom=2mm, left=2mm, right=2mm, boxsep=0mm, breakable}}
\newtcolorbox[auto counter]{Frame}[1][]{%
	enhanced,
	breakable,
	colback=white,
	colbacktitle=white,
	coltitle=black,
	fonttitle=\bfseries,
	before skip=6pt,
	after skip=6pt,
	boxrule=1.25pt,
	titlerule=.2pt,
	toptitle=1.5pt,
	bottomtitle=1.5pt,
	title=Frame~\thetcbcounter,
	#1}
\definecolor{lightyellow}{RGB}{255, 255, 204}
\sethlcolor{lightyellow}

\usepackage[utf8]{inputenc}
\usepackage[T1]{fontenc}

\usepackage[american]{babel}

\usepackage[right=1in, left=1in, top=1in, bottom=1in]{geometry}
\usepackage{pdflscape}
\usepackage{amsmath}
\usepackage{amssymb}
\usepackage{amsfonts}
\usepackage{bm}
\usepackage{bbm}
\usepackage{graphicx}
\usepackage{xcolor}
\usepackage{colortbl}
\usepackage[figuresright]{rotating}
\usepackage[capposition=top]{floatrow}

\usepackage{caption}
\usepackage{subcaption}
\DeclareCaptionFormat{cont}{#1 (cont.)#2#3\par}
\usepackage{array}
\usepackage{booktabs}
\usepackage{multirow}
\usepackage{tabularx}
\usepackage{longtable}
\usepackage{adjustbox}
\usepackage[flushleft]{threeparttable}
\usepackage{dcolumn}
\usepackage{makecell}
\usepackage{ragged2e}
\usepackage{siunitx}
\usepackage{enumitem}

\RequirePackage[colorlinks,citecolor=blue,urlcolor=blue,breaklinks]{hyperref}
\usepackage{xurl}

\usepackage[title]{appendix}

\usepackage[gen]{eurosym}
\usepackage{authblk}
\usepackage{setspace}
\usepackage{comment}
\usepackage{footmisc}
\usepackage{numprint}
\usepackage{datetime}
\usepackage{lipsum} 
\usepackage{color}

\usepackage{xpatch} 
\usepackage[%
      style=authoryear,%
      maxcitenames=2,%
      maxbibnames=5,%
      backend=biber,		
      natbib=true,
      uniquename=init,
      giveninits=true,
      url=true,
      uniquelist=false 
	]{biblatex}
\usepackage{csquotes}

\renewbibmacro{in:}{%
  \ifentrytype{article}{}{\printtext{\bibstring{in}\intitlepunct}}}

\DeclareFieldFormat{citehyperref}{%
  \DeclareFieldAlias{bibhyperref}{noformat}
  \bibhyperref{#1}}

\DeclareFieldFormat{textcitehyperref}{%
  \DeclareFieldAlias{bibhyperref}{noformat}
  \bibhyperref{%
    #1%
    \ifbool{cbx:parens}
      {\bibcloseparen\global\boolfalse{cbx:parens}}
      {}}}

\savebibmacro{cite}
\savebibmacro{textcite}

\renewbibmacro*{cite}{%
  \printtext[citehyperref]{%
    \restorebibmacro{cite}%
    \usebibmacro{cite}}}

\renewbibmacro*{textcite}{%
  \ifboolexpr{
    ( not test {\iffieldundef{prenote}} and
      test {\ifnumequal{\value{citecount}}{1}} )
    or
    ( not test {\iffieldundef{postnote}} and
      test {\ifnumequal{\value{citecount}}{\value{citetotal}}} )
  }
    {\DeclareFieldAlias{textcitehyperref}{noformat}}
    {}%
  \printtext[textcitehyperref]{%
    \restorebibmacro{textcite}%
    \usebibmacro{textcite}}}

\DeclareFieldFormat[online]{url}{%
  \begingroup
  \footnotesize  
  \url{#1}%
  \endgroup
} 

\DeclareFieldFormat[article]{doi}{%
  \href{https://dx.doi.org/#1}{\nolinkurl{#1}}}
\DeclareFieldFormat[book]{doi}{%
  \href{https://dx.doi.org/#1}{\nolinkurl{#1}}}
\DeclareFieldFormat[inproceedings]{doi}{%
  \href{https://dx.doi.org/#1}{\nolinkurl{#1}}}

\AtEveryBibitem{%
  \clearfield{month}%
  \clearfield{day}%
  \clearfield{issn}%
  \clearfield{isbn}%
  \ifentrytype{article}{%
    \clearfield{url}%
    \clearfield{urldate}%
    \clearfield{publisher}%
    \clearfield{series}%
    \clearfield{note}%
  }{}%
  \ifentrytype{book}{%
    \clearfield{url}%
    \clearfield{urldate}%
    \clearfield{series}%
    \clearfield{note}%
  }{}%
  \ifentrytype{inproceedings}{%
    \clearfield{url}%
    \clearfield{urldate}%
    \clearfield{series}%
    \clearfield{note}%
    \clearfield{publisher}%
    \clearfield{location}%
  }{}%
}

\xpatchbibmacro{volume+number+eid}{%
  \setunit*{\adddot}%
}{%
}{}{}
\DeclareFieldFormat[article]{number}{\mkbibparens{#1}}

\xpatchbibmacro{date+extrayear}{
  \printtext[parens]%
}{%
  \setunit{\addcomma\space}%
  \printtext%
}{}{}

\DeclareFieldFormat{pages}{#1}
\title{%
  \LARGE\bfseries
  AI Innovation and Firm Performance in the \\ Medical Device Industry\\[0.6em]%
}
\author[1,2]{Fazliddin Shermatov\thanks{%
  Corresponding author: \texttt{shermatov@unistra.fr}}}
\author[3]{Stéphane Robin}
\author[4,5]{Aldo Geuna}
\affil[1]{Université Sorbonne Paris Nord, France}
\affil[2]{Université de Strasbourg, France}
\affil[3]{Université Paris~1 Panthéon-Sorbonne, France}
\affil[4]{Department of Cultures, Politics and Society, University of Turin, Italy}
\affil[5]{Collegio Carlo Alberto, Turin, Italy}
\date{\monthname[\the\month]~\the\year}

\begin{document}
\maketitle

\begin{center}
  \vspace{-1em}
  \textit{Working Paper}
  \vspace{0.5em}
\end{center}

{%
  \let\thefootnote\relax
  \footnotetext{\small The authors declare no conflicts of interest.
  The views expressed are those of the authors and do not necessarily
  reflect those of any affiliated institution.}%
}

\begin{abstract}
\begin{onehalfspacing}
\noindent Whether artificial intelligence pays off for the firms that build it into their products is hard to establish, because AI innovation is itself hard to observe. The medical technology sector is a rare exception: an AI-enabled device must obtain clearance from a national health authority before it can reach a patient, leaving a dated, firm-attributable record of AI innovation output that can be observed directly rather than proxied. We exploit this setting with a three-stage recursive model estimated on a novel firm-level dataset linking FDA premarket clearances, USPTO patents, Scopus publications, and Orbis financials, tracing the full innovation chain from external collaboration through AI device introduction to firm performance. We find that external AI research collaboration is a robust driver of AI device introduction across firm sizes and estimators, with a larger effect for small firms, consistent with external knowledge ties substituting for limited internal R\&D capacity. Decomposing by partner type, the effect is largest for industry and clinical collaborations and smallest for academic ties, consistent with the former being closer to the regulatory and commercialisation process. Firms that bring AI devices to market display higher labour productivity — an effect robust for small firms and the full sample that holds under both sequential and joint maximum-likelihood estimation and accumulates across successive device introductions. Effects on profit margins are present but weaker and do not survive all specifications, a pattern consistent with competitive entry eroding pricing power as AI devices diffuse through the sector.

\vspace{1em}

\vspace{0.4em}
\end{onehalfspacing}
\end{abstract}

\clearpage
\doublespacing

\section{Introduction}

Whether artificial intelligence pays off for the firms that enhance their products with it is difficult to establish, because AI innovation is itself difficult to observe. Firm-level studies typically proxy it with the demand for AI skills in job postings, resume-based measures of AI hiring, or patent counts — measures of intent or input that rarely correspond to a product that has actually reached a market \parencite{Alekseeva2020, Babina2024, Alderucci2020}. The medical technology (hereafter, medtech) sector is a rare exception. An AI-enabled medical device must obtain clearance from a national health authority before it can reach a patient, leaving a dated, firm-attributable record of the AI innovation output itself. This setting lets us observe not merely that a firm invests in AI, but that it has succeeded in bringing a regulated AI-enhanced product to market.

The healthcare AI sector is among the fastest-growing segments of the global economy, estimated at roughly \$37bn in 2025, projected above \$744bn by 2035 \parencite{Precedence2026}. Applications of AI to medical and health-related technologies include medical imaging, natural language processing for clinical documentation, and predictive analytics to count a few. With North America capturing the largest share of investment \parencite{Landi2025, HTR2025}, strategic acquisitions such as Tempus AI's \$81.25 million purchase of Paige and its \$600 million acquisition of Ambry Genetics illustrate an accelerating trend toward platform consolidation and multimodal data integration \parencite{Tempus2025, Tempus2025b}. Yet how individual medtech firms build the capabilities behind these products, and what those capabilities deliver in terms of commercial outcomes, remains largely unexamined. This paper aims to fill this gap addressing two related questions: \textit{How do medical technology firms develop capabilities to innovate in AI-enhanced medical devices?} And: \textit{How does AI device innovation affect firm performance?}

To answer these questions, we construct a novel firm-level dataset that links four sources: FDA premarket clearance records for both conventional and AI-enabled medical devices, patent data from the USPTO via PatentsView, scientific publication records from Scopus, and financial data from Orbis. The dataset covers firms active in three medical specialties: Radiology, Cardiovascular Pathologies, and Neurology which together account for more than 90\% of all FDA-cleared AI devices. Crucially, it spans the full innovation chain from capability-building inputs (publications and external collaboration) through innovation outputs (AI device approvals) to market outcomes (labour productivity and profit margins). This design, which directly connect AI inputs to regulated innovation outputs, distinguishes our work from most existing empirical studies of AI and firm performance, which rely either on survey data, resume-based measures of AI investment, or financial market outcomes \parencite{Babina2024, Seamans2018}.

We estimate a three-stage recursive model. In the first stage, external AI collaboration intensity is instrumented by the firm's AI publication count and its geographic distance to the nearest AI research hub. In the second stage, fitted collaboration drives the introduction of AI device(s) on the market, measured both as a binary outcome and as a cumulative device count, and instrumented in both cases by the flow of AI-specific patents, constructed from the Artificial Intelligence Patent Dataset \parencite[AIPD;][]{Pairolero2025}. In the third stage, the predicted AI device introduction is linked to firm performance. This recursive structure allows us to isolate the performance effect attributable to the collaboration-driven component of AI innovation, rather than to unobserved firm heterogeneity.

We find that collaboration in research on AI significantly increases the probability of introduction of AI device(s) and that this effect is larger for small firms, consistent with external knowledge ties substituting for limited internal R\&D capacity. Decomposing by partner type, the marginal effect of collaboration is largest for industry and clinical partners and smallest for academic collaboration, with the former more closely connected to the regulatory validation and commercialisation processes needed for device clearance. Second, the introduction of AI devices raises labour productivity robustly, but its effect on profit margins is weaker and does not survive all specifications. Productivity gains persist because they are embedded in the firm's production process, whereas margin gains appear to erode. This suggests that as AI devices diffuse through the sector, competitive entry is eroding pricing power. The productivity premium is robust across estimation approaches and estimators, holding under both sequential and joint maximum-likelihood estimation.

The rest of the paper is organised as follows. We review the relevant literature in Section~\ref{sec:lit}. We describe the data in Section~\ref{sec:data} and set out the empirical strategy in Section~\ref{sec:strategy}. We present the results in Section~\ref{sec:results} and conclude in Section~\ref{sec:conclusion}.

\section{Literature Review}
\label{sec:lit}

There is a broad case for treating AI as a general-purpose technology \parencite{Agrawal2019, Crafts2021} or a general method of invention \parencite{Bianchini2022, Cockburn2018}. Regardless of framing, its impact on science and innovation is considerable, and it is expected to reshape a wide range of sectors, from manufacturing and finance to healthcare \parencite{Davenport2018}. Early empirical evidence points to meaningful efficiency gains at the task and occupation level \parencite{Brynjolfsson2025}, while broader reviews suggest more mixed aggregate effects alongside wide-ranging consequences for labour markets, income distribution, and governance \parencite{Comunale2024, AcemogluRestrepo2020}.

\subsection{Innovation and firm performance}
\label{sec:inovfirm}

A large body of evidence establishes that innovation raises firm performance, with effects varying by innovation type, firm size, and competitive conditions \parencite{HallLottiMairesse2009}. Process innovations reduce costs and raise labour productivity; product innovations expand revenues through differentiation and demand creation \parencite{Griffith2006, CirilloMina2023}. The CDM framework \parencite{Crepon1998, LoofMairesseMohnen2017, NottenMairesseVerspagen2017} provides the canonical econometric structure for linking these stages, modelling innovation inputs, outputs, and productivity as a recursive system that addresses the endogeneity at each step \parencite{Hall2011}. 

Most empirical work on the innovation--productivity link has relied on survey-based measures of innovation inputs and outputs, such as those from Community Innovation Surveys \parencite{Crepon1998, Griffith2006, HallMairesse2010, Hall2011, MairesseRobin2012, MairesseRobin2017}. Early productivity accounting showed that conventional factor inputs --- capital and labour --- explained less than half of observed output growth, with the residual attributed to technical change; subsequent work has sought to connect that residual explicitly to measurable knowledge investments \parencite{Griliches1996, Hall2011}. More recent studies using administrative and market data have extended this agenda. Robot adoption, measured from International Federation of Robotics shipment data, has been linked to substantial productivity growth at the country and industry level \parencite{GraetzMichaels2015}.Among manufacturing firms, digital technology adoption raises labour productivity and is associated with human capital upscaling, consistent with this broader pattern \parencite{CetteNevoux2022, Moncada2025}.

For AI specifically, the evidence is more mixed and the identification problem more acute. Fewer than 6\% of US firms reported any use of AI-related technologies in 2018, with adoption concentrated in large, young, and technology-intensive establishments and geographically clustered in a small number of innovation hubs \parencite{McElheran2024}. At the European level, prior ICT capability and knowledge complementarities in computing and network technologies are the primary determinants of whether firms innovate in AI at all \parencite{IgnaVenturini2023}. This concentration complicates inference on performance effects. When AI adoption is proxied by demand for AI skills in job postings, associations with labour productivity are not robust, though management-oriented AI adoption is positively related to revenue and investment growth \parencite{Alekseeva2020}. 

Using patent-based measures linked to US Census microdata, \textcite{Alderucci2020} find positive productivity and labour demand effects. \textcite{Babina2024}, using resume-based measures of AI hiring, find that AI-investing firms grow faster through product innovation, with returns concentrated among larger firms with
complementary resources. Yet firm-level evidence from France indicates that the productivity premium among AI users largely reflects the selection of already-productive firms into adoption rather than treatment effects per se, with the exception of firms that develop AI in-house \parencite{CalvinoFontanelli2026}. \textcite{Seamans2018} distil this into a methodological requirement: clean identification demands granular firm-level data that directly connects AI inputs to regulated innovation outputs and performance, rather than self-reports or indirect proxies.

\subsection{AI in healthcare}
\label{sec:aihealth}

In healthcare specifically, AI is widely expected to reshape medicine by improving the experience of both clinicians and patients \parencite{Topol2019, Rajpurkar2022}. Empirical progress has been most visible in medical imaging: deep-learning models have matched or exceeded clinical specialists in interpreting chest radiographs, retinal scans, dermatological images, and electrocardiograms, while natural language processing tools have reduced the burden of clinical documentation, and risk-stratification models have begun to improve patient triage and discharge planning \parencite{Topol2019, Rajpurkar2022, ObermeierEmanuel2016}. The medical device sector sits at the commercial end of this pipeline: innovations developed in academic and clinical AI research are eventually packaged into regulated products that must obtain regulatory clearance before reaching patients.

Reviews of AI adoption in healthcare institutions identify a consistent set of enabling and inhibiting factors \parencite{Roppelt2024}. On the enabling side, the volume of untapped health data is considerable: it has been estimated that 97\% of health data assets go unused \parencite{Thomason2021}, creating potential for AI tools to extract diagnostic and operational value. On the inhibiting side, an algorithm is only as good as the data used to train it \parencite{Singh2020}; training-data artefacts can embed systematic biases that degrade performance in deployment populations, a concern documented empirically in predictive health algorithms \parencite{ObermeierEmanuel2016}. AI model performance can additionally degrade over time as patient populations, protocols, or infrastructure change \parencite{Pianykh2020}. Infrastructure requirements are non-trivial: high-performance computing, high-speed networks, and robust data storage are preconditions for effective AI deployment \parencite{Noorbakhsh2019}. Transparency about algorithmic development and clinical involvement in model design has been identified as important for adoption \parencite{Watson2020}.

The regulatory environment shapes both the pace and direction of AI device development. In the United States, the FDA evaluates medical devices through three main pathways: the premarket approval (PMA) pathway for high-risk devices, the de~novo review for low-to-moderate risk novel devices, and the 510(k) premarket notification pathway, which grants clearance to devices substantially equivalent to a predicate already on the market. Empirical analyses of FDA authorisations confirm that the large majority of AI-enabled devices have entered the market through the 510(k) pathway, and that Radiology and Cardiology account for the bulk of approvals \parencite{Benjamens2020, Apell2021}. In the European Union, a parallel architecture of regulation governs medical devices (MDR), in-vitro diagnostics (IVDR), health technology assessment (HTAR), and health data access (EHDS), to which the AI Act of 2024 adds horizontal requirements for high-risk AI systems covering transparency, risk management, and lifecycle accountability \parencite{AIAct2024, EUCouncil2024MDR}. These regulatory structures affect firms' innovation strategies and their ability to appropriate returns, since the cost and timeline of regulatory compliance are non-trivial fixed costs that differ by device type and risk class.

Improvements in medical technology have historically been a primary driver of increased life expectancy and rising health expenditure \parencite{Newhouse1992, Cutler2004}. The economics of medical innovation give a central role to market exclusivity: without the temporary market power created by patents and regulatory data exclusivity, firms would underinvest in R\&D \parencite{Nordhaus1969}. A substantial empirical literature confirms that enlarged market opportunities drive R\&D investment in healthcare \parencite{WardDranove1995, AccemogluLinn2004, Finkelstein2004}. AI, however, may alter the standard logic in important ways. Unlike conventional process innovations that lower marginal costs and trigger competitive adoption, AI adoption does not necessarily generate negative externalities for non-adopters that compel market-wide uptake \parencite{AIdiffusion2024}. Data assets also create persistent advantages for incumbents, since proprietary patient data is not easily replicable by entrants \parencite{Eisfeldt2023}.

\subsection{Knowledge collaboration and firm innovation}
\label{sec:collab}

Firms increasingly rely on external knowledge to generate innovations, and absorptive capacity has become a central determinant of how effectively they do so \parencite{Cohen1990, Arora2018}. Knowledge from universities, public research organisations, hospitals, and other firms all constitute important components of firms' external knowledge flows \parencite{Cohen2002, Arvanitis2008, Caloghirou2021, Tether2008, RobinSchubert2013, GonzalezPernia2015}. Crucially, the breadth of engagement across these diverse external sources -i.e. the openness of the firm- predicts innovation performance, and not merely the depth of any single relationship \parencite{Laursen2006}. These collaborations occur through diverse channels, including joint publications and co-patenting, R\&D alliances and contracts, consulting, technology licensing, and the mobility of trained scientists and engineers \parencite{Perkmann2007, Perkmann2013, BodasFreitas2013, Hagedoorn2002}.

Different partner types provide qualitatively distinct knowledge inputs. Collaboration with universities and public research organisations supplies frontier scientific knowledge that is inherently tacit and embedded in scientific practice, and therefore difficult to transfer through codified channels alone \parencite{Perkmann2007, VegaJurado2017}. Collaboration with other firms is particularly suited to accessing complementary production capabilities and process knowledge that neither side could develop alone, and tends to involve more codified technology transfer \parencite{Hagedoorn2002, Miotti2003}. This reflects a broader make-or-buy logic in R\&D: firms with sufficient absorptive capacity treat external and internal R\&D as complements, while those with weaker internal capabilities tend to substitute external sourcing for in-house development \parencite{CassVeug2006}. In the medtech industry, a third distinct input comes from clinical partners, who supply domain expertise and access to patient data that neither pure research institutions nor commercial firms typically have internally \parencite{Caloghirou2021, GonzalezPernia2015}. In technology-intensive sectors, the complementarity between these knowledge streams means that firms engaging with a broader range of partner types access knowledge that no single partner type can supply alone \parencite{Belderbos2004, Laursen2006}. This complementarity is especially pronounced in the medical device industry, where development simultaneously demands, on the one hand, engineering and algorithmic capability and, on the other, clinical knowledge - ingredients that are rarely present within a single organisation.

The empirical evidence on the innovation effects of external knowledge collaboration is broadly positive, though the magnitude depends on firm characteristics and institutional context \parencite{YuLee2017, Szucs2018, VegaJurado2017, GarciaVega2020, AnonHigon2016}. Collaboration with research partners tends to be more impactful for new-to-market product introductions than for incremental improvements, consistent with research partners supplying frontier rather than codified knowledge \parencite{BargeGil2019}. Firm size is a key mediating factor. Smaller firms face tighter constraints on internal R\&D budgets and are less likely to employ specialised scientific staff capable of independently tracking a fast-moving research frontier \parencite{Veugelers1997, Arora2018}. \textcite{Belderbos2004} find that the innovation productivity gain from collaboration with universities is concentrated among smaller, less R\&D-intensive firms in technology-intensive industries. For firms with limited internal capability, external collaboration can therefore substitute for lacking in-house scientific capacity rather than complement it \parencite{Veugelers1997, Fudickar2019}.

Geographic proximity to external knowledge sources reinforces these effects through localised knowledge spillovers. Innovative activity clusters within a short radius of universities and public laboratories, since transmitting tacit knowledge across distance is difficult \parencite{Jaffe1989, AudretschFeldman1996, Arundel2004}. Proximity lowers the costs of forming collaborative partnerships and facilitates repeated, informal interactions through which scientific knowledge is shared most productively. In biotechnology, \textcite{ZuckerDarby1998} provide evidence that firms physically co-located with academic scientists were substantially more innovative, in part because proximity was necessary to internalise tacit knowledge embodied in the cutting-edge field. In the pharmaceutical industry, \textcite{CockburnHenderson1998} find that firms with active co-authorship relationships with academic scientists were significantly more productive in drug discovery. Both findings motivate the identification strategy in Section~\ref{sec:strategy}, where geographic distance to AI research hubs and a firm's stock of AI-related scientific publications serve as excluded instruments for its external collaboration intensity.

In the AI context, the dependence of industrial innovation on external knowledge is particularly high. The foundational methods of modern AI were developed predominantly within academic computer science and statistics, and the research frontier continues to advance through the public scientific literature and open-source code/databases \parencite{Cockburn2018, Bianchini2022}. \textcite{AroraGambardella1990} find that, in biotechnology, from the late 1970s, academic science became a direct input into industrial R\&D, raising the returns to collaboration for firms positioned to exploit it. This dependence is compounded for regulated AI-enabled medical devices. Translating an algorithmic method into an FDA-cleared device demands the simultaneous mobilisation of machine-learning expertise and clinical knowledge. Very few firms can sustain this interdisciplinary combination without relying on inputs that universities, hospitals, and research networks supply together, rather than any single partner type in isolation \parencite{Topol2019, Rajpurkar2022}. The empirical implication is that firms with stronger ties to the AI research community should face lower effective costs of device development and therefore be better positioned to introduce AI-enhanced products to market. We return to this distinction empirically in Section~\ref{sec:strategy} and ~\ref{app:robustness}. There we decompose external collaboration intensity by partner type, academic, company, healthcare, and other, to assess whether the commercialisation-relevant channels identified above are also those driving AI device introduction.

\section{Data}
\label{sec:data}

\subsection{AI-Enabled Medical Devices}

We use the FDA's official list of Artificial Intelligence-Enabled Medical Devices to capture AI innovation in the medical device sector. The list was retrieved in December 2025 and contained 1,247 approved devices, of which approximately 94\% received clearance through the 510(k) premarket notification pathway. We filter specialties with more than 50 registered devices in order to focus on medical device sub-sectors significantly impacted by AI. Three specialties account for over 90\% of all AI device approvals: Radiology (956 devices), Cardiovascular (116), and Neurology (56).

We retrieve all 510(k) premarket approvals from the FDA database, yielding 125,678 devices as of December 2025. This step helps us to situate AI adoption within the broader medical device landscape and create a control group. Within the three specialties above, 30,886 devices were filed by 7,290 distinct applicants. Our analysis covers all companies with at least one premarket-approved medical device in any of the three aforementioned specialties.

\subsection{Financial Data}

We match applicant company names to financial records via Orbis, a global company database maintained by Bureau van Dijk (BvD) that provides harmonised balance sheet, income statement, and ownership data across jurisdictions. The matching yields 5,652 unique BvD identifiers. The match rate (78\%) reflects three main sources of attrition: variant name spellings that resolve to a single BvD identifier after de-duplication; recently incorporated firms not yet carrying financial history on Orbis; and a residual share of companies simply absent from the database. For each matched company we compile annual financial data over the period 2000--2025, including revenues, total assets, R\&D expenditure, employment, profitability ratios, and capital structure variables.

The United States accounts for the large majority of the sample, with 4,057 companies (72\% of the total). European Economic Area countries and the United Kingdom together constitute the second-largest group, with 784 companies (14\% of the total). The Asia-Pacific region accounts for a further 593 companies (11\%), led by China including Hong Kong (261 companies). The remaining companies are distributed across a further 26 countries, each contributing fewer than 25 observations. Fewer than 10\% of companies with non-missing employment data are large companies with more than 500 employees. 

\subsection{Publication Data}

We retrieve scientific publication records via affiliation search on Scopus, querying the full publication history of the 5,652 companies in our sample. The search yields 1,062,119 distinct publications as of January 2026. To identify AI-relevant output, we cross-reference publication identifiers against the corpus of 5.4 million AI papers compiled by \textcite{Bianchini2025AIAS}, which covers AI-related publications from 1960 onward and is constructed from a curated set of keyword-based filters applied to Scopus metadata.

Publication activity has grown steadily over the sample period. Of the 1,062,119 total publications, the bulk fall within the period since the early 2010s. Annual output in that period averages approximately 35,000 papers per year, rising to 52,361 by 2025. This corresponds to a compound annual growth rate (CAGR) of over 2\% since 2010. Restricting to AI-related output, publications date back to the 1960s. Of the 72,560 firm-affiliated AI-related papers, nearly three-quarters (72\%) were published after 2012.

For each company-year we construct the following bibliometric measures: the total number of AI publications and the average number of unique institutional affiliations per AI paper as a measure of collaboration breadth. Our key variable of interest in Stage~1 is the average number of AI-related external collaborations. It is constructed as follows. For each firm-year, we retain only those publications identified as AI-related via the \textcite{Bianchini2025AIAS} corpus. For each AI publication, we use the Scopus author metadata to count the number of distinct external institutional affiliations among co-authors, excluding the focal firm's own affiliation. An affiliation is classified as external if it does not match the firm's registered name or any of its known subsidiaries and alternate spellings identified during the de-duplication step.

Finally, we average this count across all AI-related publications authored by the firm in a given year. This yields a continuous measure of collaboration intensity that captures both the breadth and regularity of the firm's engagement with the external AI research community. For example, a firm that co-authors a single AI paper with 1 external partner in a given year receives a score of 1; a firm that co-authors 3 AI papers, each involving an average of 4 distinct external partners, receives a score of 4. Therefore, this measure reflects the firm's embeddedness in the scientific AI network rather than just its publication volume. Years in which a firm has no AI-related publications are assigned a value of zero.

To examine which type of external partner accounts for this aggregate measure, we further classify each external affiliation by partner type using the Research Organization Registry (ROR) typology. ROR types \textit{Education} and \textit{Facility} are grouped into an \textit{academia} category, covering universities and large-scale research infrastructures. \textit{Healthcare} affiliations are classified as \textit{healthcare}, and \textit{Company} affiliations as \textit{company}. All remaining ROR types (Government, Funder, Nonprofit, Archive, and Other) and any unclassified affiliations form a residual \textit{others} category. For each AI publication, we count the number of distinct external affiliations falling into each category, again excluding the firm itself, and average these category-specific counts across a firm's AI publications in a given year. This yields four partner-type collaboration measures, $academia\_collab$, $company\_collab$, $healthcare\_collab$, and $others\_collab$, which sum to the aggregate measure described above.
 
Academic affiliations account for the largest share of the 26,271 classified external affiliations (37.7\%), followed by others (27.3\%), healthcare (17.7\%), and company affiliations (17.3\%). At the publication level: 81.4\% of AI publications include at least one academic co-affiliation, 34.9\% include at least one company co-affiliation, and 21.2\% include at least one healthcare co-affiliation. Academic ties are therefore the most pervasive form of external collaboration, while company and healthcare partnerships are more narrowly targeted. We acknowledge that the others category might include miss-classified company research centers.

\subsection{Patent Data}

We retrieve patent records from PatentsView, which draws on the USPTO patent grant and application databases, yielding 1,387,219 patents across the companies in our sample as of January 2026. For each company, we construct 3 patent-based indicators: a knowledge stock, a patent flow, and an AI-specific patent flow.

The \textit{knowledge stock} accumulates all patents granted up to and including year $t$. To account for the economic obsolescence of knowledge assets over time, we apply the perpetual inventory method with annual depreciation rate $\delta$:
\begin{equation}
\label{eq:depreciation}
    S_{it} = P_{it} + (1 - \delta)\, S_{i,t-1}
\end{equation}
\noindent where $S_{it}$ is the knowledge stock of firm $i$ in year $t$, $P_{it}$ is the number of patents granted in that year, and $\delta \in \{0.10; 0.15\}$ in accordance with the common practice in the literature. The stock is initialised from the earliest observed patent year for each firm so that pre-sample patenting activity is reflected at the start of the observation window. We use a depreciation rate of 15\% as the baseline, consistent with rates applied to technology-intensive sectors. We also experimented with an alternative depreciation rate of 10\%, which did not significantly affect our results.

The \textit{patent flow} records the number of patents filed by the firm in a given year. Filing dates are used to closely reflect the timing of underlying innovation activity.

The \textit{AI patent flow} ($PAF_{it}$) measures AI-specific patenting activity at the firm-year level. Each patent is matched to the Artificial Intelligence Patent Dataset \parencite{Pairolero2025}, which covers USPTO patent documents published through 2023, and provides machine-learning-based probability scores indicating whether a patent belongs to an AI or machine learning technology class. We retain patents in the highest-likelihood AI/ML tier and compute, for each firm-year, the count of AI patents per year, which yields $PAF_{it}$.

\begin{figure}[t]
  \centering
  \includegraphics[width=\textwidth]{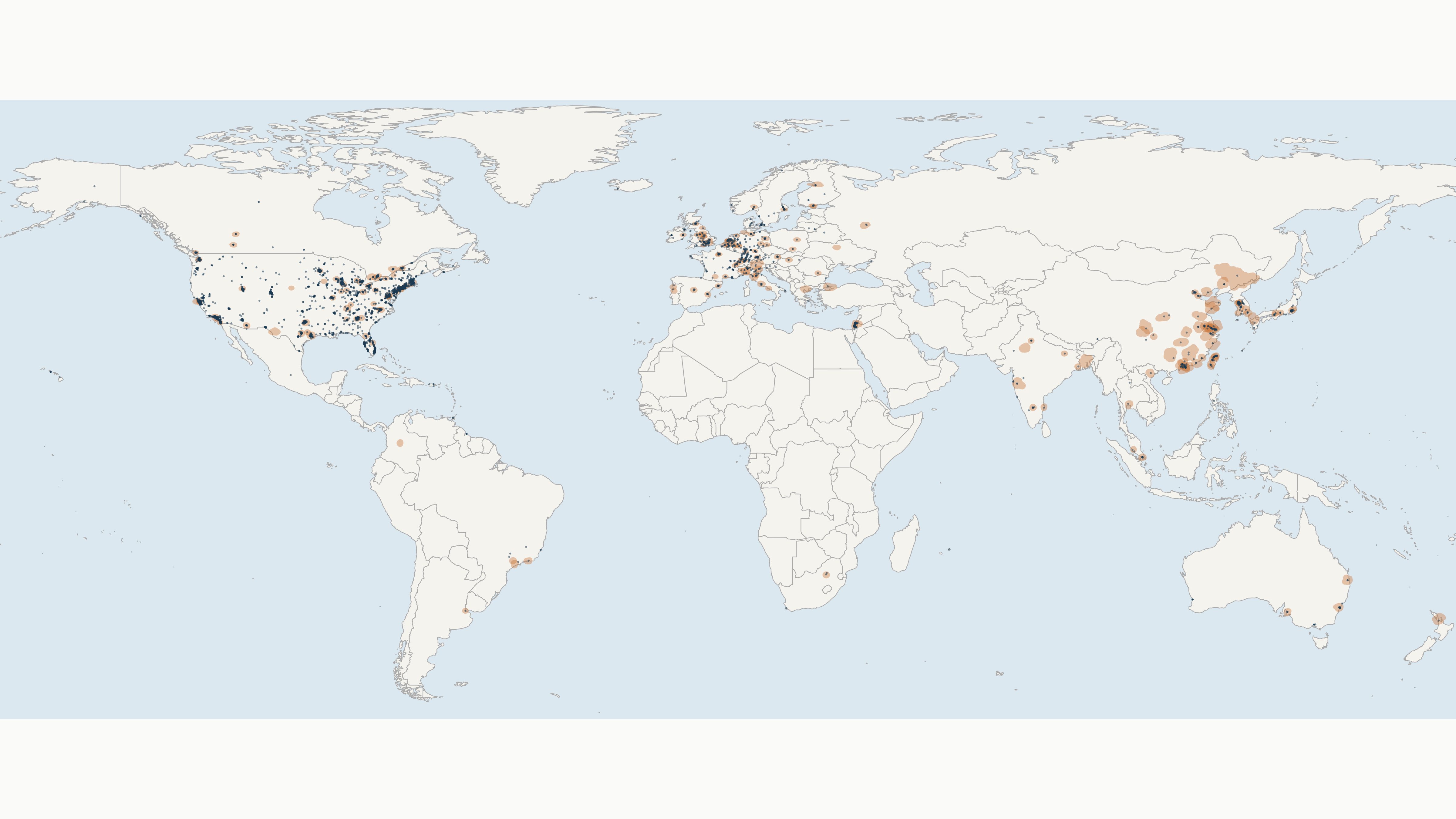}
  \caption{\textbf{Geographic Distribution of FDA-Registered Medical Device Firms and AI Research Hubs.}}
  \floatfoot{\textit{Note:} Each dot represents a unique firm location drawn from the FDA 510(k) clearance database. Shaded regions (80\,km radius) denote cities with concentrations of high-impact AI research output.}
  \label{fig:fda_ai_hubs_map}
\end{figure}

\subsection{Distance to AI Hubs}

To capture proximity to knowledge externalities in artificial intelligence, we construct a geographic distance measure for each company. Using the \textcite{Bianchini2025AIAS} corpus of 5.4 million AI publications, we identify 300 AI hotspot cities based on publication volume normalised by citation intensity. For each company in our sample we then compute the geodesic distance in kilometres from its registered address to the nearest hotspot city. This yields a time-varying instrument, as the set of hotspot cities is recomputed annually, so that a company's distance reflects the evolving geography of AI research rather than a fixed baseline.

\begin{table}[t]
\centering
\caption{Regression sample descriptive statistics}
\label{tab:desc_stats}
\setlength{\tabcolsep}{6pt}
\renewcommand{\arraystretch}{1.05}
\begin{adjustbox}{max width=\linewidth}
\small
\begin{tabular}{lrrrrrr}
\toprule
 & \textbf{N} & \textbf{Mean} & \textbf{SD} & \textbf{Min} & \textbf{Median} & \textbf{Max} \\
\midrule
\rowcolor[gray]{0.95}
AI device dummy                    & 9322 & \num{0.048}       & \num{0.214}        & \num{0.000}   & \num{0.000}   & \num{1.000}          \\
AI devices (cumulated)             & 9322 & \num{0.126}       & \num{1.110}        & \num{0.000}   & \num{0.000}   & \num{64.000}         \\
\rowcolor[gray]{0.95}
Profit margin                       & 5068  & \num{1.917}      & \num{26.637}       & \num{-211.743}   & \num{5.987}  & \num{98.980}        \\
Value added per employee (th.\ USD) & 2976  & \num{132.103}     & \num{281.188}      & \num{0.002}   & \num{81.340}  & \num{7991.636}       \\
\rowcolor[gray]{0.95}
Collab.\ in AI research (avg)      & 9322 & \num{0.325}       & \num{1.480}        & \num{0.000}   & \num{0.000}   & \num{47.000}         \\
Capital, $K$ (th.\ USD)             & 5806  & \num{272595.800}  & \num{2788095.900}  & \num{0.000}   & \num{509.189} & \num{65797000.000}   \\
\rowcolor[gray]{0.95}
Missing $K$ dummy                  & 9322 & \num{0.377}       & \num{0.485}        & \num{0.000}   & \num{0.000}   & \num{1.000}          \\
Employees, $L$                     & 9322 & \num{5131.120}    & \num{28636.530}    & \num{0.000}   & \num{35.000}  & \num{440000.000}     \\
\rowcolor[gray]{0.95}
AI patent flow (PAF)              & 9322 & \num{4.354}      & \num{32.091}      & \num{0.000}   & \num{0.000}   & \num{615.000}       \\
Patent stock                        & 9322 & \num{238.328}     & \num{1561.352}     & \num{0.000}   & \num{0.354}   & \num{22261.090}      \\
\rowcolor[gray]{0.95}
Start-up dummy                     & 9322 & \num{0.049}       & \num{0.215}        & \num{0.000}   & \num{0.000}   & \num{1.000}          \\
New entrant dummy                   & 9322 & \num{0.218}       & \num{0.413}        & \num{0.000}   & \num{0.000}   & \num{1.000}          \\
\rowcolor[gray]{0.95}
Distance to AI hub (km)            & 8774 & \num{838.629}      & \num{1547.651}      & \num{0.000}   & \num{76.482}  & \num{7021.875}       \\
AI publications                     & 9322 & \num{0.856}       & \num{15.269}       & \num{0.000}   & \num{0.000}   & \num{1328.000}       \\
\bottomrule
\end{tabular}
\end{adjustbox}
\begin{minipage}{\linewidth}
\smallskip\footnotesize
\textit{Notes:} Sample restricted to firm-year observations from 2012 onwards. Variables are in levels; logged versions are constructed in the regressions. Capital and value added per employee are expressed in thousands of USD. Missing $K$ dummy equals 1 when capital is unobserved.
AI PAF\,=\,AI patent application flow. Distance to AI hub is the distance (km) to the nearest AI research centre.
\end{minipage}
\end{table}

\subsection{Start-Ups and New Entrants}

We identify two categories of companies that require special treatment given their distinct financial profiles. Start-ups are defined as companies with fewer than 10 employees whose age at observation does not exceed 5 years since their recorded incorporation date, where this information is available. Age is computed from the incorporation date variable in the Orbis records. Companies satisfying both the size and age criteria in a given year are flagged with a binary start-up indicator; 366 unique companies receive this flag. Start-ups are of particular interest in this context, as early-stage ventures are disproportionately likely to enter the medical device market with AI-enabled products.

New entrants are companies that first appear with non-missing financial data (defined as having at least one valid observation for employees, value added, capital, profit margin, revenue, or intangibles) only in the last year of the sample (2025), but are not classified as start-ups. This group comprises 2,098 companies and consists of established firms whose financials became available on Orbis with a delay rather than genuinely young ventures. A separate binary indicator is constructed for this group to control for the mechanical correlation between recent Orbis coverage and measured financial performance, which would otherwise bias the estimation of firm-level outcomes in Stage~3.

\section{Empirical Analysis}
\label{sec:strategy}

Our empirical analysis relies on the previously-mentioned CDM framework \parencite{Crepon1998}, adapted to the specific medtech AI context. The innovation process is modelled as a recursive system of econometric equations (without feedback). The system is recursive in the sense that the equations are nested in three clearly-defined stages, each stage being modelled as the determinant of the next one: Firms build AI capability through external collaborations (Stage~1); AI capability drives the possibility to introduce AI-enhanced devices on the market (Stage~2); as a form of (AI-enabled) product innovation, the introduction of such devices may affect firm profitability and performance (Stage~3). We address endogeneity in this sequential structure by including instruments in the first and second stage.

The 3-stage model can be formalized as follows:

\begin{equation}
\label{eq:CDM}
\left\{
\begin{aligned}
&y_{1it} = \mathbf{x}_{it}'\beta_1 + \mathbf{z}_{1it}'\gamma_1 + \mathbf{z}_{2it}'\gamma_2 + \epsilon_{1it} \\[0.4em]
&y_{2it} = g\left(\alpha_2 \cdot y_{1it} + \mathbf{x}_{it}'\beta_2 + \mathbf{z}_{2it}'\gamma_2 + \epsilon_{2it} \right) \\[0.4em]
&\ln y_{3it} = \alpha_3 \cdot y_{2it} + \mathbf{x}_{it}'\beta_3 + \epsilon_{3it}
\end{aligned}
\right.
\end{equation}

where $y_{1it}$ is the average number of AI-related external collaborations in which firm $i$ is involved in year $t$, $y_{2it}$ is a measure of success in AI-related product innovation, $y_{3it}$ is a measure of firm performance (profit margin or labour productivity), $\mathbf{x}_{it}$ are regressors common to all three stages and $\mathbf{z}_{1it}$ and $\mathbf{z}_{2it}$ are vectors of instruments. Vector $\mathbf{z}_{1it}$ is specific to the first stage, whereas $\mathbf{z}_{2it}$ is included in both the first and second stages. It operates as a second-stage instrument (with respect to the third stage). 

The first and third stages are always specified as linear equations, since $y_{1it}$ and $y_{3it}$ are always continuous variables. The specification of the second stage, however, depends on the nature of $y_{2it}$, which may either (i) be a dummy variable indicating that firm $i$ has introduced at least one AI-enhanced medical device on the market over the observation period, or (ii) measure the number of devices introduced over the observation period. When $y_{2it}$ is a dummy variable, then $g(.)$ can either be specified as linear (giving rise to the Linear Probability Model, hereafter LPM) or be equal to the indicator function ($g(.) = \mathbbm{1}\!\left(.\right)$), which gives rise to the Probit model. When  $y_{2it}$ measures the number of AI-enhanced devices that firm $i$ has introduced, then $g(.)$ is specified as a Tobit model.

Controls~$\mathbf{x}_{it}$ include the log-stock of capital $\ln K$, log-employment $\ln L$, log-patenting intensity $\ln P/I$ (where P is the stock of patents computed using the perpetual inventory method and where I is firm revenue), as well as country, industry and year fixed effects.\footnote{Technically, in order to handle zeros, we define the log-stock of capital as $\ln(1+K)$, log-employment as $\ln(1+L)$ and log-patenting intensity as $\ln (1+Patents)/(1+Revenue)$. For the sake of convenience, though, we adopt the simpler notations $\ln K$, $\ln L$ and  $\ln P/I$ in the remainder of this paper.}
Information on the stock of capital is missing for a great many firms. We do not want to exclude these firms from our sample, since their other characteristics are observed and we fear that excluding them would entail important information loss. We do want to keep the capital variable as a control variable, though. In order to solve this conundrum, we include a "missing information on capital stock" dummy variable in our regressors ($No~Capital$). We conduct two sensitivity analyses, first estimating Model~\eqref{eq:CDM} without the $No~Capital$ dummy, and second estimating it without any control for capital stock. We present the results of these sensitivity analyses in  \ref{app:robustness}. These results do not modify the conclusions of our main estimations regarding the key links between AI capability building, innovation in AI and firm performance.
Last but not least, we do not include individual fixed effects because the panel is extremely unbalanced and many firms appear only once in the sample. This means that we cannot control for individual unobserved heterogeneity (UH), but in CDM-type models such as Model~\eqref{eq:CDM} the primary concern is endogeneity. We therefore make every effort to effectively control for endogeneity in our framework, using an instrumental variable (IV) approach as our basic tool and Pooled 2SLS as our benchmark estimator (see \textcite{SeWo2010} for the merits of Pooled 2SLS in estimating econometric models on panel data).

Since Model~\eqref{eq:CDM} comprises three nested stages, we need two sets of instruments, one set for the first stage and one for the second stage. In the above, we have denoted these two sets $\mathbf{z}_{1it}$ and $\mathbf{z}_{2it}$, respectively. In $\mathbf{z}_{1it}$ we consider two instruments candidates: (i) $ln(Pub_{it})$, the log count of the firm’s AI-related scientific publications, and (ii) $Distance$, the geodesic distance to the nearest quality AI research hub. The rationale is that both factors can shift the intensity of external collaboration without plausibly affecting through any direct channel the registration (and introduction on the market) of AI-enhanced medical devices, nor firm performance. In $\mathbf{z}_{2it}$ we find a single instrument: the log-flow of AI-related patents applied for ($\ln PAF_{it}$) lagged two years. This provides a viable instrument insofar as its effect on profitability and productivity operates solely through the AI innovation channel. Ideally, additional instrument candidates would strengthen our analysis, possibly allowing for overidentification. One interesting possibility would be variables that could explain collaboration in AI while being related to the firm's partners rather than to the firm itself, such as a measure (or indicator) of public funding / grants received for research in AI by a collaborating university lab. We currently do not have such variables at our disposal, but it is a path we want to explore in our future research.

\subsection{Estimation strategy: benchmark analysis}
\label{Benchmark}

In our benchmark analysis, our measure of AI product innovation, $y_{2it}$,is a dummy variable indicating whether firm $i$ has introduced at least one AI-enhanced medical device on the market\footnote{Devices are considered as introduced once they have received clearance from the national health authority.} at time $t$. This leads us to estimate two alternative specifications of Model~\eqref{eq:CDM}. In the first specification, we assume $y_{2it}$ is a linear function of the second-stage regressors and instruments, i.e. we specify the second stage as a LPM. This allows us to estimate the first two stages of Model~\eqref{eq:CDM} by Pooled 2SLS as an IV regression :

\begin{subequations}
\label{eq:dummywhy2}
\renewcommand{\theequation}{\theparentequation.\alph{equation}}

\begin{equation}
\label{eq:2SLS}
\left\{
\begin{aligned}
y_{1it} &= \mathbf{x}_{it}'\beta_1 + \mathbf{z}_{1it}'\gamma_1 + \mathbf{z}_{2it}'\gamma_2 + \epsilon_{1it} \\[0.4em]
y_{2it} &= \alpha_2 \cdot y_{1it} + \mathbf{x}_{it}'\beta_2 + \mathbf{z}_{2it}'\gamma_2 + \epsilon_{2it} \\[0.4em]
\end{aligned}
\right.
\end{equation}

\smallskip

In the second specification, we assume $y_{2it}$ is linked to the second-stage regressors and instruments by the indicator function $\mathbbm{1}\!\left(.\right)$, i.e. we specify the second stage as a Probit model. This allows us to estimate the first two stages of Model~\eqref{eq:CDM} by Maximum Likelihood (ML) as an IV Probit model :

\begin{equation}
\label{eq:IVProbit}
\left\{
\begin{aligned}
y_{1it} &= \mathbf{x}_{it}'\beta_1 + \mathbf{z}_{1it}'\gamma_1 + \mathbf{z}_{2it}'\gamma_2 + \epsilon_{1it} \\[0.4em]
y_{2it} &= 
\mathbbm{1}\!\left(\alpha_2 \cdot y_{1it} + \mathbf{x}_{it}'\beta_2 + \mathbf{z}_{2it}'\gamma_2 + \epsilon_{2it} > 0  \right) \\[0.4em]
\end{aligned}
\right.
\end{equation}

\end{subequations}

\smallskip

\noindent As a sensitivity analysis, we also estimate Model~\eqref{eq:IVProbit} using \citet{New1987}'s minimum chi-square two-step estimator. The results, presented in \ref{app:robustness}, are qualitatively quite similar to those obtained by ML. 

From Model~\eqref{eq:2SLS}, we derive $\hat{y}_{2it}$, the predicted \textit{linear} probability that firm $i$ has introduced an AI-enhanced medical device at time $t$. This predicted probability is then used as a regressor in the third stage of Model~\eqref{eq:CDM}, with $\mathbf{z}_{2it}$ acting as an instrument in the second stage:

\begin{subequations}
\label{eq:3rdStage}
\renewcommand{\theequation}{\theparentequation.\alph{equation}}

\begin{equation}
\label{eq:linear}
\ln y_{3it} = \alpha_3 \cdot \hat{y}_{2it} + \mathbf{x}_{it}'\beta_3 + \epsilon_{3it}
\end{equation}

\noindent From Model~\eqref{eq:IVProbit}, we derive $\hat{p}_{2it} = Prob(y_{2it}=1)$, the predicted probability that firm $i$ has introduced an AI-enhanced medical device at time $t$. This predicted probability is then used as a regressor in the third stage of Model~\eqref{eq:CDM}, again with $\mathbf{z}_{2it}$ acting as an instrument in the second stage:

\begin{equation}
\label{eq:prob}
\ln y_{3it} = \alpha_3 \cdot \hat{p}_{2it} + \mathbf{x}_{it}'\beta_3 + \epsilon_{3it}
\end{equation}

\end{subequations}

\noindent We estimate Equations \eqref{eq:linear} and \eqref{eq:prob} by OLS, using bootstrapped standard errors (with 300 replications) to control for predicted regressor bias. 

To complement our benchmark analysis, we decompose our measure of collaborations in AI  $y_{1it}$ by type of partner. As described in Section~\ref{sec:data} we define, based on ROR organisation type, four types of partner: academic, company, healthcare and other. We re-estimate Model~\eqref{eq:2SLS} within each category by 2SLS, using the same set of instruments. We report the results of these complementary regressions in ~\ref{app:robustness}, Table~\ref{tab:collab_decomp}. The decomposition is intended to identify which type of external partner may drive the benchmark result. It does not replace the main specification that involves the global measure of external collaborations in AI.

\subsection{Estimation strategy: Tobit extension}
\label{Tobit}

The models estimated in Sub-Section \ref{Benchmark} capture only the extensive margin of AI innovation, i.e. whether firms manage to introduce at least one AI-enhanced medical device. We now try to go further and capture the intensive margin of AI innovation, exploiting the fact that our data also contains information on the number of devices introduced every year between 2012 and 2025. Keeping in mind that, for a majority of firms, this number is either 0 or 1, we now define $y_{2it}$ as the total number of AI-enhanced medical devices registered and brought to the market by firm $i$ at time $t$. Since this variable is left-censored at zero, we now specify the second stage of Model~\eqref{eq:CDM} as a Tobit model. This allows us to estimate the first two stages of Model~\eqref{eq:CDM} by ML) as an IV Tobit model: 

\begin{equation}
\label{eq:IVtobit}
\left\{
\begin{aligned}
&y_{1it} = \mathbf{x}_{it}'\beta_1 + \mathbf{z}_{1it}'\gamma_1 + \mathbf{z}_{2it}'\gamma_2 + \epsilon_{1it} \\[0.4em]
&\mathbb{E}(y_{2it} \mid y_{1it}, \mathbf{x}_{it}, \mathbf{z}_{2it}) = \Phi\!\left(\frac{\alpha_2 \cdot y_{1it} + \mathbf{x}_{it}'\beta_2 + \mathbf{z}_{2it}'\gamma_2}{\sigma_2}\right)
\left(\alpha_2 \cdot y_{1it} + \mathbf{x}_{it}'\beta_2 + \mathbf{z}_{2it}'\gamma_2 + \sigma_2 \lambda_2\right)
\end{aligned}
\right.
\end{equation}

\smallskip 

\noindent where $\lambda_2$ denotes the inverse Mills ratio and where $\sigma_2$ denotes the standard deviation of $y_{2it}$. 

From Model \eqref{eq:IVtobit}, we derive $\hat{y}_{2it}$, the predicted number of AI-enhanced medical devices that firm $i$ has brought to the market at time $t$. This predicted number is then used as a regressor in the third stage of Model~\eqref{eq:CDM}, with $z_{2it}$ acting as an instrument in the second stage:

\begin{equation}
\label{eq:posttobit}
\ln y_{3it} = \alpha_3 \cdot \hat{y}_{2it} + \mathbf{x}_{it}'\beta_3 + \epsilon_{3it}
\end{equation}

\noindent As before, we estimate Equation \eqref{eq:posttobit} by OLS, using bootstrapped standard errors (with 300 replications) to control for predicted regressor bias

As an alternative to the above estimation procedure, we also estimate all three stages of Model~\eqref{eq:CDM}, with its second stage specified as an IV Tobit, by Full Information Maximum Likelihood (FIML) using Stata's Conditional Mixed Process (CMP) procedure \parencite{Roodman2011}:

\begin{equation}
\label{eq:CDM2}
\left\{
\begin{aligned}
&y_{1it} = \mathbf{x}_{it}'\beta_1 + \mathbf{z}_{1it}'\gamma_1 + \mathbf{z}_{2it}'\gamma_2 + \epsilon_{1it} \\[0.4em]
&\mathbb{E}(y_{2it} \mid y_{1it}, \mathbf{x}_{it}, \mathbf{z}_{2it}) = \Phi\!\left(\frac{\alpha_2 \cdot y_{1it} + \mathbf{x}_{it}'\beta_2 + \mathbf{z}_{2it}'\gamma_2}{\sigma_2}\right)
\left(\alpha_2 \cdot y_{1it} + \mathbf{x}_{it}'\beta_2 + \mathbf{z}_{2it}'\gamma_2 + \sigma_2 \lambda_2\right) \\[0.4em]
&\ln y_{3it} = \alpha_3 \cdot y_{2it} + \mathbf{x}_{it}'\beta_3 + \epsilon_{3it}
\end{aligned}
\right.
\end{equation}

\smallskip

An important difference between this approach and the sequential approach we used previously is that, with FIML estimation, we never use the predicted value of the outcome of a given stage as a regressor in the next stage. Instead, as can be seen from Equation~\eqref{eq:CDM2}, the second and third stages include the \textit{actual} value of the outcome of the previous stage among the regressors. Model~\eqref{eq:CDM2} also takes into account cross-equation error correlation. The correlation coefficients of the errors enter its Likelihood as parameters to be estimated. In practice, the CMP procedure takes the estimates of each individual equation in Model~\eqref{eq:CDM2} as initial values for the Likelihood of the full model. The Likelihood is then maximized using a numerical algorithm. 

\section{Results}
\label{sec:results}

We estimate all the models laid out in Section~\ref{sec:strategy} on three samples: (1) small firms with 500 employees or fewer, (2) large firms, and (3) the full sample.

\subsection{Benchmark estimates}
\label{BenchmarkResults}

\subsubsection{AI product innovation: Extensive margin}
\label{BenchmarkStageII}

\begin{table}[htbp]
\centering
\caption{Benchmark 2-stages estimates}
\label{tab:benchmark_2stages}
\setlength{\tabcolsep}{4pt}
\renewcommand{\arraystretch}{1.05}
\begin{adjustbox}{max width=\linewidth}
\footnotesize
\begin{tabular}{>{\raggedright\arraybackslash}p{5.2cm}cccccc}
\toprule
 & \multicolumn{2}{c}{\textbf{Small firms}}
 & \multicolumn{2}{c}{\textbf{Large firms}}
 & \multicolumn{2}{c}{\textbf{All firms}} \\
\cmidrule(lr){2-3}\cmidrule(lr){4-5}\cmidrule(lr){6-7}
 & \textbf{(1) 2SLS} & \textbf{(2) IV probit}
 & \textbf{(3) 2SLS} & \textbf{(4) IV probit}
 & \textbf{(5) 2SLS} & \textbf{(6) IV probit} \\
\midrule
\multicolumn{7}{l}{\textit{2\textsuperscript{nd} Stage — Outcome: AI dummy (introduction of $\geq$1 AI device)}} \\[2pt]
\rowcolor[gray]{0.95}
Collab. in AI research (avg)
  & $0.047^{***}$ & $0.209^{***}$
  & $0.028^{***}$ & $0.109^{**}$
  & $0.031^{***}$ & $0.165^{***}$ \\
\rowcolor[gray]{0.95}
  & $(0.008)$ & $(0.029)$ & $(0.009)$ & $(0.045)$ & $(0.006)$ & $(0.023)$ \\
Log $K$
  & $-0.0004$ & $-0.003$
  & $-0.005^{**}$ & $-0.087^{***}$
  & $-0.003^{***}$ & $-0.026^{**}$ \\
  & $(0.001)$ & $(0.017)$ & $(0.002)$ & $(0.023)$ & $(0.001)$ & $(0.012)$ \\
\rowcolor[gray]{0.95}
Missing $K$ dummy
  & $-0.059$ & $-0.620$
  & $-0.449^{*}$ & $-8.094^{***}$
  & $-0.279^{**}$ & $-2.931^{**}$ \\
\rowcolor[gray]{0.95}
  & $(0.131)$ & $(1.705)$ & $(0.260)$ & $(2.422)$ & $(0.109)$ & $(1.227)$ \\
Log $L$
  & $0.008^{***}$ & $0.098^{***}$
  & $0.001$ & $0.182^{**}$
  & $0.003^{*}$ & $0.085^{***}$ \\
  & $(0.002)$ & $(0.025)$ & $(0.005)$ & $(0.087)$ & $(0.002)$ & $(0.018)$ \\
\rowcolor[gray]{0.95}
Log AI P.A.F.
  & $0.004^{**}$ & $0.021^{**}$
  & $0.005^{**}$ & $0.068^{***}$
  & $0.002^{*}$ & $0.018^{**}$ \\
\rowcolor[gray]{0.95}
  & $(0.002)$ & $(0.010)$ & $(0.001)$ & $(0.021)$ & $(0.001)$ & $(0.008)$ \\
Log Patenting
  & $0.002^{***}$ & $0.033^{***}$
  & $-0.007^{***}$ & $-0.086^{***}$
  & $0.001^{***}$ & $0.017^{***}$ \\
  & $(0.0005)$ & $(0.010)$ & $(0.001)$ & $(0.017)$ & $(0.0005)$ & $(0.005)$ \\
\rowcolor[gray]{0.95}
Start-up
  & $0.024^{**}$ & $0.331^{***}$
  & $\bar{\phantom{x}}$ & $\bar{\phantom{x}}$
  & $0.016$ & $0.301^{***}$ \\
\rowcolor[gray]{0.95}
  & $(0.012)$ & $(0.118)$ & & & $(0.011)$ & $(0.112)$ \\
New entrant
  & $0.002$ & $0.074$
  & $-0.173^{*}$ & $-0.235^{*}$
  & $-0.004$ & $0.024$ \\
  & $(0.009)$ & $(0.099)$ & $(0.090)$ & $(0.690)$ & $(0.010)$ & $(0.094)$ \\
\rowcolor[gray]{0.95}
Country FE ($p$-val.)
  & $0.169$ & $0.001$ & $0.001$ & $0.535$ & $0.442$ & $0.021$ \\
Industry FE ($p$-val.)
  & $0.000$ & $0.000$ & $0.000$ & $0.000$ & $0.000$ & $0.000$ \\
\rowcolor[gray]{0.95}
Year FE ($p$-val.)
  & $0.000$ & $0.000$ & $0.000$ & $0.000$ & $0.000$ & $0.000$ \\[3pt]
\midrule
\multicolumn{7}{l}{\textit{1\textsuperscript{st} Stage — Outcome: Collaborations in AI research (avg.\ number)}} \\[2pt]
\rowcolor[gray]{0.95}
Log $K$
  & $0.003$ & $0.003$ & $0.017$ & $0.044^{**}$ & $0.005$ & $0.009^{*}$ \\
\rowcolor[gray]{0.95}
  & $(0.005)$ & $(0.005)$ & $(0.016)$ & $(0.019)$ & $(0.006)$ & $(0.005)$ \\
Missing $K$ dummy
  & $0.329$ & $0.329$ & $1.436$ & $4.252^{**}$ & $0.439$ & $0.825$ \\
  & $(0.529)$ & $(0.527)$ & $(1.723)$ & $(1.972)$ & $(0.605)$ & $(0.527)$ \\
\rowcolor[gray]{0.95}
Log $L$
  & $0.033^{***}$ & $0.033^{***}$
  & $-0.022$ & $-0.097$
  & $0.018^{**}$ & $0.019^{***}$ \\
\rowcolor[gray]{0.95}
  & $(0.009)$ & $(0.009)$ & $(0.056)$ & $(0.069)$ & $(0.008)$ & $(0.007)$ \\
Log AI P.A.F.
  & $-0.007$ & $-0.007$ & $0.013$ & $-0.011$ & $-0.003$ & $-0.009$ \\
  & $(0.007)$ & $(0.007)$ & $(0.011)$ & $(0.013)$ & $(0.006)$ & $(0.006)$ \\
\rowcolor[gray]{0.95}
Log Patenting
  & $0.004^{**}$ & $0.004^{**}$
  & $0.021^{***}$ & $0.015^{**}$
  & $0.008^{***}$ & $0.005^{***}$ \\
\rowcolor[gray]{0.95}
  & $(0.002)$ & $(0.002)$ & $(0.007)$ & $(0.007)$ & $(0.002)$ & $(0.002)$ \\
Start-up
  & $0.036$ & $0.036$
  & $\bar{\phantom{x}}$ & $\bar{\phantom{x}}$
  & $0.023$ & $0.024$ \\
  & $(0.038)$ & $(0.038)$ & & & $(0.041)$ & $(0.038)$ \\
\rowcolor[gray]{0.95}
New entrant
  & $-0.026$ & $-0.026$
  & $0.954$ & $1.190$
  & $-0.015$ & $-0.015$ \\
\rowcolor[gray]{0.95}
  & $(0.025)$ & $(0.025)$ & $(0.720)$ & $(0.964)$ & $(0.025)$ & $(0.025)$ \\
Distance from nearest AI hub (km)
  & $-0.00003$ & $-0.00002$
  & $-0.0001$ & $0.0001$
  & $-0.00002$ & $-0.00001$ \\
  & $(0.00002)$ & $(0.00002)$ & $(0.0001)$ & $(0.0001)$ & $(0.00002)$ & $(0.00002)$ \\
\rowcolor[gray]{0.95}
Log AI publications
  & $2.878^{***}$ & $2.879^{***}$
  & $1.438^{***}$ & $2.282^{***}$
  & $1.827^{***}$ & $2.596^{***}$ \\
\rowcolor[gray]{0.95}
  & $(0.269)$ & $(0.269)$ & $(0.100)$ & $(0.296)$ & $(0.095)$ & $(0.203)$ \\
Country FE ($p$-val.)
  & $0.002$ & $0.002$ & $0.560$ & $0.539$ & $0.000$ & $0.010$ \\
Industry FE ($p$-val.)
  & $0.244$ & $0.241$ & $0.000$ & $0.075$ & $0.418$ & $0.303$ \\
\rowcolor[gray]{0.95}
Year FE ($p$-val.)
  & $0.012$ & $0.007$ & $0.057$ & $0.457$ & $0.002$ & $0.002$ \\
\midrule
$\text{atanh}\,\rho$
  & $\bar{\phantom{x}}$ & $-0.146^{***}$
  & $\bar{\phantom{x}}$ & $-0.065$
  & $\bar{\phantom{x}}$ & $-0.112^{***}$ \\
  & & $(0.038)$ & & $(0.068)$ & & $(0.031)$ \\
\rowcolor[gray]{0.95}
$\text{Log}\,\sigma$
  & $\bar{\phantom{x}}$ & $-0.186$
  & $\bar{\phantom{x}}$ & $0.276^{**}$
  & $\bar{\phantom{x}}$ & $-0.088$ \\
\rowcolor[gray]{0.95}
  & & $(0.125)$ & & $(0.113)$ & & $(0.095)$ \\
\midrule
Observations & $7{,}131$ & $7{,}131$ & $1{,}643$ & $1{,}093$ & $8{,}774$ & $8{,}429$ \\
Firms        & $3{,}912$ & $3{,}912$ &   $234$ &   $166$ & $4{,}094$ & $4{,}063$ \\
\bottomrule
\end{tabular}
\end{adjustbox}
\begin{minipage}{\linewidth}
\smallskip \scriptsize
\textit{Notes:} $^{***}$ $p<0.01$, $^{**}$ $p<0.05$, $^{*}$ $p<0.1$. Clustered robust standard errors in parentheses. 

FE rows report $p$-values of $F$-tests ($H_0$: FE parameters are all zero). 2SLS global $F$-tests are significant at 1\% in both stages. The ``start-up'' dummy variable is omitted for large firms since it is always equal to zero in this sub-sample. IV Probit models are estimated by ML; $\rho$ is the correlation coefficient of the first- and second-stage errors; $\sigma$ is the standard deviation of the first-stage error. Wald tests reject exogeneity ($\text{atanh}\,\rho=0$) at 1\% for small firms and all firms, confirming endogeneity of ``collaborations in AI''. Marginal effects of ``avg.\ collaborations in AI'' at sample mean: $0.012^{***}$ $(0.002)$ for small firms; $0.006^{**}$ $(0.003)$ for large firms; $0.011^{***}$ $(0.002)$ for all firms. The sample size for large firms (and, to a lesser extent, for all firms) drops slightly when estimating IV Probits due to perfect prediction for some country-industry-year. 
\end{minipage}
\end{table}

Table~\ref{tab:benchmark_2stages} reports the results of Models \eqref{eq:2SLS} and \eqref{eq:IVProbit}. Collaborations in AI research have a statistically significant positive impact on the introduction of a new AI-enhanced medical device across all three samples. Namely, according to the LPM 2SLS estimates, a one-unit increase in predicted AI collaboration intensity raises the probability of AI device introduction by 4.7 percentage points for small firms, 2.8 percentage points for large firms, and 3.1 percentage points for the full sample. The IV Probit confirms this pattern. At sample mean, an additional AI collaborator raises the probability of device entry by 1.2 percentage points for small firms, 0.6 percentage points for large firms, and 1.1 percentage points for the full sample. 

Among control variables, firm size in terms of employment (Log $L$) is positively associated with the introduction of AI device(s), which is consistent with larger firms possessing greater absorptive capacity to convert external knowledge into commercialized AI devices. The AI patent application flow (Log AI P.A.F.) enters positively and significantly across all samples, which concurs with the view that firms with stronger pre-existing AI technological capabilities are better positioned to introduce AI-enhanced products.

The start-up dummy is positive and significant under the IV Probit specification, for small firms (0.331) and the full sample (0.301). Young firms are therefore more likely to introduce an AI-enhanced device, conditional on their AI collaboration intensity. Several explanations are plausible. Start-ups may face fewer organisational and legacy-technology constraints than incumbents. Alternatively, some start-ups may be built around an AI-enabled product from the entry. The coefficient is not estimable for large firms, as none qualify as start-ups. As we discuss in Section~\ref{sec:conclusion}, this higher propensity to introduce AI devices does not translate into a productivity premium for start-ups in the third stage.


\smallskip

\begin{table}[htbp]
\centering
\caption{Post-estimation tests for 2SLS estimations}
\label{tab:postest_2sls}
\setlength{\tabcolsep}{12pt}
\renewcommand{\arraystretch}{1.1}
\begin{adjustbox}{max width=\linewidth}
\small
\begin{tabular}{lccc}
\toprule
 & \textbf{Small firms} & \textbf{Large firms} & \textbf{All firms} \\
 & \textbf{(1)} & \textbf{(2)} & \textbf{(3)} \\
\midrule
\multicolumn{4}{l}{\textbf{Underidentification Test} \textit{(H$_0$: Equation is underidentified)}} \\
~~Kleibergen-Paap LM statistic & $152.79$ & $263.59$ & $455.58$ \\
~~$p$-value                    & $0.000$  & $0.000$  & $0.000$  \\
\midrule
\multicolumn{4}{l}{\textbf{Weak Identification Test} \textit{(H$_0$: Instruments are weakly correlated)}} \\
~~Kleibergen-Paap $F$ statistic & $59.29$  & $104.28$ & $186.78$ \\
~~Critical value               & $19.93$  & $19.93$  & $19.93$  \\
\midrule
\multicolumn{4}{l}{\textbf{Overidentification Test} \textit{(H$_0$: Instruments are valid/exogenous)}} \\
~~Hansen $J$ statistic         & $1.615$  & $2.265$  & $5.854$  \\
~~$p$-value                    & $0.204$  & $0.132$  & $0.016$  \\
\bottomrule
\end{tabular}
\end{adjustbox}
\begin{minipage}{\linewidth}
\smallskip\footnotesize
\textit{Notes:} The underidentification test utilizes the Kleibergen-Paap LM statistic. The weak identification test reports the Kleibergen-Paap Wald $F$ statistic against the Stock-Yogo critical value of 10\%, which ensures that a test at the 5\% level of significance cannot have a worst-case rejection probability above 10\%.  The overidentification test is performed with the heteroskedasticity-robust Hansen $J$ statistic.
\end{minipage}
\end{table}

\paragraph{Instrument performance.}
The instruments perform well across all samples. Kleibergen--Paap $F$-statistics of 59.3, 104.3, and 186.8 for small firms, large firms, and the full sample respectively rule out weak instruments throughout. The Hansen~$J$ statistic does not reject overidentification for small firms ($J=1.6$, $p=0.204$) nor large firms ($J=2.3$, $p=0.132$); the full-sample test rejects overidentification at the 5\% level ($J=5.9$, $p=0.016$).

Log AI publications enters the first stage with large, highly significant coefficients ($2.878^{***}$, $1.438^{***}$, and $1.827^{***}$ for small, large, and all firms respectively in the 2SLS estimates), confirming strong relevance. The exclusion restriction is satisfied because a firm's accumulated AI scientific output does not directly predict device registration or commercialisation. It shapes those outcomes only through the collaboration channel it enables, and is accordingly excluded from the second stage. Geodesic distance to the nearest AI hub provides a complementary source of exogenous variation in collaboration costs, and together the two instruments deliver the strong first-stage performance documented above.

The validity of the instrument set rests importantly on the properties of firm-level AI publications. This variable captures a firm's scientific capacity in artificial intelligence, which raises its attractiveness as a collaboration partner and lowers the search and matching costs of forming AI-related R\&D partnerships. This  satisfies the relevance condition, as confirmed by the strong first-stage $F$-statistics well above the Stock--Yogo critical value of 19.93.  The exclusion restriction holds because a body of published research does not itself constitute a regulatory submission or generate commercial revenue: AI publications affect device clearance and commercialisation only insofar as that scientific capacity is translated into product development through actual collaborative R\&D. The geodesic distance to major AI research hubs provides a complementary source of exogenous variation in collaboration costs. Together, the two instruments are sufficiently powerful and pass the overidentification test cleanly across the two sub-samples. 

\paragraph{Partner-type decomposition.} Appendix Table~\ref{tab:collab_decomp} compares the 2SLS estimates for all firms presented in Table~\ref{tab:benchmark_2stages} with the 2SLS estimates obtained when Model \eqref{eq:2SLS} is estimated separately for each type of partner. In all four categories (Academia, Company, Healthcare and Others), the effect of collaboration in AI research on the introduction of AI device(s) remains positive and strongly significant. Academic collaboration is both the most prevalent form of external tie, present in over 80\% of AI publications (Section~\ref{sec:data}), and the most strongly instrumented, with a first-stage Kleibergen--Paap $F$-statistic of 1,759.5. Company collaboration carries the largest point estimate (0.355) despite being the rarest tie, present in around 35\% of AI publications and accounting for only 17.3\% of classified affiliations, representing a narrow but commercially proximate channel. The coefficients rise in magnitude from academia to healthcare to company. This ordering broadly tracks the rarity and commercialisation-proximity of each tie. We retain the aggregate measure as our headline specification because it captures the firm's overall collaboration intensity across all affiliation types, while the partner-type decomposition serves as a robustness and mechanism check.

\subsubsection{Effect of introducing an AI-enhanced medical device on firm performance}
\label{BenchmarkStageIII}

\begin{table}[htbp]
\centering
\caption{Benchmark third stage linear regressions}
\label{tab:benchmark_3rdstage}
\setlength{\tabcolsep}{4pt}
\renewcommand{\arraystretch}{1.05}
\begin{adjustbox}{max width=\linewidth}
\small
\begin{tabular}{>{\raggedright\arraybackslash}p{5.2cm}cccccc}
\toprule
 & \multicolumn{2}{c}{\textbf{Small firms}}
 & \multicolumn{2}{c}{\textbf{Large firms}}
 & \multicolumn{2}{c}{\textbf{All firms}} \\
\cmidrule(lr){2-3}\cmidrule(lr){4-5}\cmidrule(lr){6-7}
 & \textbf{(1) 2SLS} & \textbf{(2) IV probit}
 & \textbf{(3) 2SLS} & \textbf{(4) IV probit}
 & \textbf{(5) 2SLS} & \textbf{(6) IV probit} \\
\midrule
\multicolumn{7}{l}{\textit{Panel A — Outcome: Log Profit Margin}} \\[2pt]
\rowcolor[gray]{0.95}
Predicted AI introduction prob.
  & $0.281$ & $-0.009$
  & $0.586^{*}$ & $0.087$
  & $0.694^{**}$ & $0.250$ \\
\rowcolor[gray]{0.95}
  & $(0.367)$ & $(0.294)$ & $(0.344)$ & $(0.288)$ & $(0.293)$ & $(0.195)$ \\
Log $K$
  & $0.006$ & $0.006$
  & $0.007$ & $0.030^{***}$
  & $0.021^{***}$ & $0.020^{***}$ \\
  & $(0.009)$ & $(0.009)$ & $(0.010)$ & $(0.011)$ & $(0.006)$ & $(0.006)$ \\
\rowcolor[gray]{0.95}
Missing $K$ dummy
  & $1.162$ & $1.142$
  & $0.566$ & $3.190^{**}$
  & $2.586^{***}$ & $2.470^{***}$ \\
\rowcolor[gray]{0.95}
  & $(0.915)$ & $(0.885)$ & $(1.092)$ & $(1.249)$ & $(0.654)$ & $(0.638)$ \\
Log $L$
  & $0.008$ & $0.012$
  & $0.047^{**}$ & $0.021$
  & $-0.004$ & $-0.0002$ \\
  & $(0.024)$ & $(0.022)$ & $(0.020)$ & $(0.023)$ & $(0.010)$ & $(0.010)$ \\
\rowcolor[gray]{0.95}
Log Patenting
  & $0.009^{*}$ & $0.010^{**}$
  & $0.007$ & $0.008$
  & $0.005$ & $0.007^{*}$ \\
\rowcolor[gray]{0.95}
  & $(0.005)$ & $(0.005)$ & $(0.006)$ & $(0.005)$ & $(0.004)$ & $(0.004)$ \\
Start-up
  & $-0.262$ & $-0.254$
  & $\bar{\phantom{x}}$ & $\bar{\phantom{x}}$
  & $-0.259^{*}$ & $-0.245$ \\
  & $(0.164)$ & $(0.167)$ & & & $(0.156)$ & $(0.160)$ \\
\rowcolor[gray]{0.95}
Country FE ($p$-val.)
  & $0.000$ & $0.000$ & $0.000$ & $0.000$ & $0.000$ & $0.000$ \\
Industry FE ($p$-val.)
  & $0.000$ & $0.000$ & $0.000$ & $0.010$ & $0.000$ & $0.000$ \\
\rowcolor[gray]{0.95}
Year FE ($p$-val.)
  & $0.102$ & $0.105$ & $0.219$ & $0.207$ & $0.126$ & $0.050$ \\
\midrule
Observations 
  & $2{,}197$ & $2{,}197$ & $1{,}497$ & $1{,}267$ & $3{,}694$ & $3{,}694$ \\
Firms 
  & $423$ & $423$ & $205$ & $169$ & $590$ & $590$ \\
\midrule
\multicolumn{7}{l}{\textit{Panel B — Outcome: Log Labour Productivity}} \\[2pt]
\rowcolor[gray]{0.95}
Predicted AI introduction prob.
  & $0.847^{**}$ & $0.984^{***}$
  & $2.024$ & $0.535$
  & $1.617^{***}$ & $1.868^{***}$ \\
\rowcolor[gray]{0.95}
  & $(0.401)$ & $(0.290)$ & $(1.156)$ & $(0.472)$ & $(0.543)$ & $(0.330)$ \\
Log $K$
  & $-0.034^{*}$ & $-0.034^{*}$
  & $-0.0003$ & $0.0002$
  & $-0.016$ & $-0.014$ \\
  & $(0.020)$ & $(0.020)$ & $(0.015)$ & $(0.019)$ & $(0.014)$ & $(0.013)$ \\
\rowcolor[gray]{0.95}
Missing $K$ dummy
  & $-8.159^{***}$ & $-8.045^{***}$
  & $\bar{\phantom{x}}$ & $\bar{\phantom{x}}$
  & $-6.260^{***}$ & $-6.064^{***}$ \\
\rowcolor[gray]{0.95}
  & $(2.681)$ & $(2.719)$ & & & $(2.208)$ & $(2.162)$ \\
Log $L$
  & $0.010$ & $0.009$
  & $0.235^{***}$ & $0.224^{***}$
  & $0.062^{***}$ & $0.047^{***}$ \\
  & $(0.032)$ & $(0.033)$ & $(0.040)$ & $(0.048)$ & $(0.021)$ & $(0.018)$ \\
\rowcolor[gray]{0.95}
Log Patenting
  & $0.026^{***}$ & $0.026^{***}$
  & $0.001$ & $-0.005$
  & $0.021^{***}$ & $0.020^{***}$ \\
\rowcolor[gray]{0.95}
  & $(0.008)$ & $(0.007)$ & $(0.009)$ & $(0.008)$ & $(0.006)$ & $(0.006)$ \\
Start-up
  & $-1.009^{***}$ & $-1.005^{***}$
  & $\bar{\phantom{x}}$ & $\bar{\phantom{x}}$
  & $-0.925^{***}$ & $-0.944^{***}$ \\
  & $(0.239)$ & $(0.246)$ & & & $(0.248)$ & $(0.238)$ \\
\rowcolor[gray]{0.95}
Country FE ($p$-val.)
  & $0.000$ & $0.000$ & $0.000$ & $0.000$ & $0.000$ & $0.000$ \\
Industry FE ($p$-val.)
  & $0.000$ & $0.000$ & $0.000$ & $0.000$ & $0.000$ & $0.000$ \\
\rowcolor[gray]{0.95}
Year FE ($p$-val.)
  & $0.227$ & $0.183$ & $0.948$ & $0.973$ & $0.012$ & $0.115$ \\
\midrule
Observations 
  & $2{,}201$ & $2{,}201$ & $775$ & $632$ & $2{,}976$ & $2{,}976$ \\
Firms 
  & $355$ & $355$ & $109$ & $85$ & $433$ & $433$ \\
\bottomrule
\end{tabular}
\end{adjustbox}
\begin{minipage}{\linewidth}
\smallskip\footnotesize
\textit{Notes:} Significance: $^{***}$ 1\%, $^{**}$ 5\%, $^{*}$ 10\%. Bootstrapped standard errors in parentheses.

FE rows report $p$-values of $F$-tests (H$_0$: FE parameters are all zero).
The Wald $\chi^2$ test of overall model significance (which replaces the usual $F$ test when standard errors are bootstrapped) rejects the null hypothesis ($p < 0.001$) across all specifications.
``New entrant'' and ``Start-up'' rows are omitted within sub-samples where their values are uniformly 0 or perfectly collinear.
In Panel B, Columns (3) and (4): the Missing $K$ dummy variable is uniformly 0 among large firms when Value Added is the outcome variable and is omitted.
\end{minipage}
\end{table}

Table~\ref{tab:benchmark_3rdstage} reports the benchmark third-stage estimates of Model \eqref{eq:CDM}, with the regressors including the predicted outcome of the second stage, i.e. either the predicted linear probability (Model \eqref{eq:linear}) or the predicted probability (Model \eqref{eq:prob}) to introduce AI device(s). We present results using the log of Value-Added per employee (i.e., the log of labour productivity) as the primary outcome and the log profit margin as an alternative outcome. The IV~Tobit and CMP estimates are reported in Sections~\ref{TobitResults} and~\ref{TobitCMP} respectively.

\paragraph{Labour productivity.}
The third-stage estimates in Panel B of Table~\ref{tab:benchmark_3rdstage} show that the predicted introduction of AI device(s) raises labour productivity across most specifications. Among small firms, the effect is positive and statistically significant using both the linear probability predicted from 2SLS and the conventional probability obtained from the IV Probit, with coefficients of $0.847$ and $0.984$ respectively. In other words, a $1\%$ increase in the probability to introduce an AI device for small firms translates to an increase in labour productivity of $0.847\%$ to $0.984\%$. For all firms, the corresponding estimates are larger in magnitude, 1.617 and 1.868 respectively, meaning that a $1\%$ increase in the probability to introduce an AI device entails a $1.617\%$ to $1.868\%$ increase in labour productivity. The large-firm coefficient is positive but insignificant in both specifications, likely reflecting the smaller sub-sample size. The consistency of the result across small firms and the full sample supports the view that the ability to introduce AI device(s) carries a significant productivity premium.

This result is in line with previous CDM-based studies, which often find a positive association between product innovation and productivity. In a conventional setting using CIS-type data in four European countries, \citet{Griffith2006} find that product innovation increases productivity by about 18\% in Spain and by about 6\% in France and the UK. \citet{HallSena2017} estimate, on 5 waves of the UK CIS, an extension of the CDM model that addresses intellectual property protection. They show that intellectual property protection bearing on product innovation raises productivity by about 12\%, and by about 15\% when the product innovation is new to the market.

Finding the full linkage from R\&D to productivity through innovation is not systematic, though. Thus, \citet{Griffith2006} do not find any significant effect of innovation (be it product or process) on productivity in the fourth country they study, Germany. Similarly, \citet{vanLeeuwenMohnen2017}, estimating a variant of CDM focused on environmental innovations on CIS data from the Netherlands, find contrasted results: Labour productivity is positively associated with resource-saving innovations and negatively associated with pollution-reducing innovations. The fact that our adaption of CDM lets us identify -- in the whole sample and for small firms -- the full linkage from collaborations in AI research to the introduction of AI devices to labour productivity is therefore noteworthy.

\paragraph{Profit margin.}
The results for profit margins, reported in Panel A of Table~\ref{tab:benchmark_3rdstage}, are weaker and less uniform than those for labour productivity. For all firms, the 2SLS estimate is positive and significant at the 5\% level ($0.694$). For large firms, the 2SLS coefficient is positive and marginally significant ($0.586$), while the IV Probit estimate for both samples is smaller and not significant statistically. Among small firms, neither specification yields a significant result. The result for large firms deserves attention: it suggests that the introduction of AI device(s) on the market may yield profit gains for established incumbents even when the evidence for small firms is absent, possibly because larger firms are better positioned to appropriate returns from AI-enabled product differentiation.

The remaining subsections complement these findings. Section~\ref{TobitResults} presents IV~Tobit estimates that exploit the intensive margin of AI adoption. Section~\ref{TobitCMP} reports the CMP joint estimator as a further robustness check.

\subsection{Tobit estimates}
\label{TobitResults}

\subsubsection{AI product innovation: Intensive margin}
\label{TobitSeq}
\begin{table}[htbp]
\centering
\caption{IV Tobit 2-stage estimates}
\label{tab:ivtobit_2stages}
\setlength{\tabcolsep}{6pt}
\renewcommand{\arraystretch}{1.1}
\begin{adjustbox}{max width=\linewidth}
\small
\begin{tabular}{>{\raggedright\arraybackslash}p{5.8cm}ccc}
\toprule
 & \textbf{Small firms} & \textbf{Large firms} & \textbf{All firms} \\
 & \textbf{(1)} & \textbf{(2)} & \textbf{(3)} \\
\midrule
\multicolumn{4}{l}{\textit{2\textsuperscript{nd} Stage (Tobit) — Outcome: Number of AI devices introduced (cumulated)}} \\[2pt]
\rowcolor[gray]{0.95}
Collaborations in AI research (avg) & $1.414^{***}$ $(0.465)$ & $1.416^{***}$ $(0.393)$ & $1.073^{***}$ $(0.317)$ \\
Log $K$                             & $-0.034$ $(0.100)$       & $-0.347^{***}$ $(0.130)$ & $-0.156^{*}$ $(0.080)$ \\
\rowcolor[gray]{0.95}
Missing $K$ dummy                   & $-5.523$ $(10.369)$      & $-28.767^{**}$ $(13.953)$ & $-17.754^{**}$ $(8.339)$ \\
Log $L$                             & $0.575^{***}$ $(0.164)$  & $-0.380$ $(0.406)$       & $0.235^{*}$ $(0.123)$ \\
\rowcolor[gray]{0.95}
Log AI P.A.F.                       & $0.170^{**}$ $(0.082)$   & $0.566^{***}$ $(0.167)$  & $0.127^{**}$ $(0.066)$ \\
Log Patenting                       & $0.187^{***}$ $(0.047)$  & $-0.787^{***}$ $(0.174)$ & $0.224^{**}$ $(0.100)$ \\
\rowcolor[gray]{0.95}
Start-up                            & $1.737^{**}$ $(0.780)$   & $\bar{\phantom{x}}$      & $1.471^{***}$ $(0.750)$ \\
New entrant                         & $0.583$ $(0.707)$        & $-10.633$ $(4.426)$      & $0.250$ $(0.682)$ \\
\rowcolor[gray]{0.95}
Country FE ($p$-val.)               & $0.030$                  & $0.000$                  & $0.333$ \\
Industry FE ($p$-val.)              & $0.000$                  & $0.000$                  & $0.000$ \\
\rowcolor[gray]{0.95}
Year FE ($p$-val.)                  & $0.022$                  & $0.027$                  & $0.000$ \\
\midrule
\multicolumn{4}{l}{\textit{1\textsuperscript{st} Stage — Outcome: Collaborations in AI research (average number)}} \\[2pt]
\rowcolor[gray]{0.95}
Log $K$                             & $0.003$ $(0.005)$        & $0.017$ $(0.016)$        & $0.005$ $(0.006)$ \\
Missing $K$ dummy                   & $0.329$ $(0.527)$        & $1.462$ $(1.703)$        & $0.435$ $(0.604)$ \\
\rowcolor[gray]{0.95}
Log $L$                             & $0.033^{***}$ $(0.009)$  & $-0.024$ $(0.054)$       & $0.018^{**}$ $(0.008)$ \\
Log AI P.A.F.                       & $-0.007$ $(0.007)$       & $0.013$ $(0.011)$        & $-0.0003$ $(0.006)$ \\
\rowcolor[gray]{0.95}
Log Patenting                       & $0.004^{**}$ $(0.002)$   & $0.021^{**}$ $(0.007)$   & $0.008^{***}$ $(0.002)$ \\
Start-up                            & $0.036^{**}$ $(0.038)$   & $\bar{\phantom{x}}$      & $0.023$ $(0.040)$ \\
\rowcolor[gray]{0.95}
New entrant                         & $-0.026$ $(0.025)$       & $0.908$ $(0.707)$        & $-0.015$ $(0.025)$ \\
Distance from nearest AI hub (km)   & $-0.00002$ $(0.00002)$   & $-0.00004$ $(0.00006)$   & $-0.00001$ $(0.00002)$ \\
\rowcolor[gray]{0.95}
Log AI publications                 & $2.878^{***}$ $(0.269)$  & $1.440^{***}$ $(0.098)$  & $1.827^{***}$ $(0.095)$ \\
Country FE ($p$-val.)               & $0.002$                  & $0.639$                  & $0.000$ \\
\rowcolor[gray]{0.95}
Industry FE ($p$-val.)              & $0.241$                  & $0.000$                  & $0.411$ \\
Year FE ($p$-val.)                  & $0.012$                  & $0.280$                  & $0.001$ \\
\midrule
\rowcolor[gray]{0.95}
$\gamma$        & $-1.264^{**}$ $(0.491)$  & $-1.439^{***}$ $(0.399)$ & $-0.841^{***}$ $(0.308)$ \\
$\log \sigma_{\epsilon_2 \vert \epsilon_1}$                            & $1.848^{***}$ $(0.195)$  & $1.999^{***}$ $(0.146)$  & $1.964^{***}$ $(0.148)$ \\
\rowcolor[gray]{0.95}
$\log \sigma_{\epsilon_1}$                            & $-0.186$ $(0.125)$       & $0.611^{***}$ $(0.189)$  & $0.122$ $(0.111)$ \\
\midrule
Observations                        & $7{,}131$                & $1{,}643$                & $8{,}774$ \\
Firms                               & $3{,}912$                & $234$                    & $4{,}094$ \\
\bottomrule
\end{tabular}
\end{adjustbox}
\begin{minipage}{\linewidth}
\smallskip\footnotesize
\textit{Notes:} Significance levels: $^{***}$ 1\%, $^{**}$ 5\%, $^{*}$ 10\%. Robust standard errors in parentheses after coefficients. 

FE rows report $p$-values of $F$-tests ($H_0$: FE parameters are jointly equal to zero). The ``start-up'' dummy variable is omitted for large firms since it is always equal to zero in this sub-sample. Parameter $\gamma$ is the partial correlation coefficient between the error terms of the second-stage (Tobit) equation and the first-stage (linear) equation, denoted $\epsilon_2$ and $\epsilon_1$ respectively: $\epsilon_{2it} = \gamma \epsilon_{1it} + \nu_{it}$, where $\nu_{it}$ is independent of $\epsilon_{1it}$ and of first- and second-stage regressors; $\log \sigma_{\epsilon_1}$ denotes the log standard deviation of $\epsilon_{1it}$, $\log \sigma_{\epsilon_2 \vert \epsilon_1}$ denotes the log standard deviation of $\epsilon_{2it}$ conditional on $\epsilon_{1it}$. Exogeneity tests systematically reject exogeneity ($H_0: \gamma = 0$), justifying the IV Tobit (rather than simple Tobit) approach. 
\end{minipage}
\end{table}

Table~\ref{tab:ivtobit_2stages} reports the IV~Tobit two-stage estimates. While the benchmark specification treats AI adoption as a binary event, the IV~Tobit exploits the full count of cumulated AI device clearances recorded in the FDA 510(k) database. This allows us to distinguish firms that introduced a single AI device from those that built a sustained portfolio of AI-enhanced products, capturing the intensive margin of AI product innovation.

The core finding from the benchmark is confirmed and sharpened. Collaborations in AI research exert a positive and highly significant effect on the cumulative number of AI devices introduced across all three subsamples. The coefficient is $1.414$ for small firms, $1.416$ for large firms, and $1.073$ for the full sample. Critically, this result holds for large firms, for whom the binary specification yielded only a marginally significant IV~Probit estimate. The Tobit framework, by drawing on variation in the depth of AI device portfolios rather than simply the entry decision, recovers a precisely estimated effect even among incumbents. This suggests that the returns to AI collaboration accumulate over successive device introductions and are not confined to the initial adoption event. 

The Wald test rejects exogeneity of AI collaborations ($H_0: \alpha = 0$) across all specifications, confirming that the IV correction remains necessary when modelling the intensive margin. The first-stage results are essentially identical to those of the benchmark. Log AI publications is the dominant instrument. Geodesic distance to the nearest AI hub retains the expected negative sign but remains imprecisely estimated, as in the benchmark. 

\begin{table}[t]
\centering
\caption{Post-IV Tobit third stage linear regressions}
\label{tab:postivtobit_3rdstage}
\setlength{\tabcolsep}{6pt}
\renewcommand{\arraystretch}{1.05}
\begin{adjustbox}{max width=\linewidth}
\small
\begin{tabular}{lcccccc}
\toprule
 & \multicolumn{2}{c}{\textbf{Small firms}} & \multicolumn{2}{c}{\textbf{Large firms}} & \multicolumn{2}{c}{\textbf{All firms}} \\
\cmidrule(lr){2-3} \cmidrule(lr){4-5} \cmidrule(lr){6-7}
 & \textbf{Profit} & \textbf{LP} & \textbf{Profit} & \textbf{LP} & \textbf{Profit} & \textbf{LP} \\
 & \textbf{(1)} & \textbf{(2)} & \textbf{(3)} & \textbf{(4)} & \textbf{(5)} & \textbf{(6)} \\
\midrule
\rowcolor[gray]{0.95}
Predicted \# of AI devices & $0.007$ & $0.029^{**}$ & $0.010^{*}$ & $0.042^{**}$ & $0.018^{**}$ & $0.047^{***}$ \\
\rowcolor[gray]{0.95}
~~(from IV Tobit)          & $(0.013)$ & $(0.013)$ & $(0.006)$ & $(0.016)$ & $(0.008)$ & $(0.016)$ \\
Log $K$                    & $0.006$ & $-0.034^{*}$ & $0.008$ & $0.002$ & $0.022^{***}$ & $-0.013$ \\
                           & $(0.008)$ & $(0.020)$ & $(0.010)$ & $(0.014)$ & $(0.006)$ & $(0.014)$ \\
\rowcolor[gray]{0.95}
Missing $K$ dummy          & $1.181$ & $-8.044^{***}$ & $0.617$ & -- & $2.695^{***}$ & $-5.928^{***}$ \\
\rowcolor[gray]{0.95}
                           & $(0.854)$ & $(2.547)$ & $(1.057)$ & & $(0.687)$ & $(2.276)$ \\
Log $L$                    & $0.007$ & $-0.000$ & $0.046^{**}$ & $0.224^{***}$ & $-0.006$ & $0.053^{***}$ \\
                           & $(0.025)$ & $(0.033)$ & $(0.020)$ & $(0.040)$ & $(0.011)$ & $(0.021)$ \\
\rowcolor[gray]{0.95}
Log Patenting              & $0.009^{*}$ & $0.023^{***}$ & $0.009$ & $0.011$ & $0.004$ & $0.018^{***}$ \\
\rowcolor[gray]{0.95}
                           & $(0.005)$ & $(0.007)$ & $(0.006)$ & $(0.010)$ & $(0.004)$ & $(0.007)$ \\
Start-up                   & $-0.267$ & $-1.041$ & -- & -- & $-0.276$ & $-0.975^{***}$ \\
                           & $(0.165)$ & $(0.252)$ & & & $(0.173)$ & $(0.249)$ \\
\rowcolor[gray]{0.95}
New entrant                & -- & -- & -- & -- & -- & -- \\
\rowcolor[gray]{0.95}
                           & & & & & & \\
\midrule
Country FE ($p$-val.)      & $0.000$ & $0.000$ & $0.000$ & $0.000$ & $0.000$ & $0.000$ \\
\rowcolor[gray]{0.95}
Industry FE ($p$-val.)     & $0.000$ & $0.000$ & $0.209$ & $0.000$ & $0.000$ & $0.000$ \\
Year FE ($p$-val.)         & $0.327$ & $0.548$ & $0.145$ & $0.776$ & $0.199$ & $0.394$ \\
\rowcolor[gray]{0.95}
Wald $\chi^2$ $p$-value    & $0.000$ & $0.000$ & $0.000$ & $0.000$ & $0.000$ & $0.000$ \\
\midrule
Observations               & $2{,}197$ & $2{,}201$ & $1{,}497$ & $775$ & $3{,}694$ & $2{,}976$ \\
Firms                      & $423$     & $355$     & $205$     & $109$ & $590$     & $433$ \\
\bottomrule
\end{tabular}
\end{adjustbox}
\begin{minipage}{\linewidth}
\smallskip\footnotesize
\textit{Notes:} Significance levels: $^{***}$ 1\%, $^{**}$ 5\%, $^{*}$ 10\%. Bootstrapped standard errors are reported in parentheses under the coefficients. Fixed Effects rows display the $p$-value of an $F$-test ($H_0$: FE parameters are jointly zero). The Wald $\chi^2$ test is reported instead of the standard $F$-test to correctly account for the properties of the bootstrapped variance-covariance matrix. In column (4), the ``Missing $K$'' dummy is omitted because it is always equal to 0 among large firms when labour productivity is the outcome variable.
\end{minipage}
\end{table}

Table~\ref{tab:postivtobit_3rdstage} reports the third-stage estimates using the predicted cumulative number of AI devices from the IV~Tobit as the endogenous regressor. This specification asks a sharper question than the benchmark: does building a deeper AI device portfolio, rather than simply crossing the adoption threshold, translate into measurable gains in firm performance?

For small firms and the full sample, the answer is affirmative. Each additional predicted AI device is associated with a $2.9$ percentage point increase in log labour productivity for small firms and a $4.7$ percentage point increase for the full sample, both statistically significant. The large-firm coefficient is positive and significant at the 5\% level in the sequential estimator ($0.042$), though this result should be interpreted with caution given that it does not survive the joint ML estimator reported in Sub-Section~\ref{TobitCMP}. The productivity premium for small firms and the full sample is robust across all specifications and estimators, and constitutes the main finding of this stage. Profit margin coefficients are positive across all sub-samples and statistically significant in the large-firm and full samples. However, as shown in Sub-Section~\ref{TobitCMP}, these effects do not remain significant under joint ML estimation.

\subsubsection{Confirmation with simultaneous estimation by ML}
\label{TobitCMP}

\begin{table}[htbp]
\centering
\caption{3-stage model, simultaneous estimation by ML (Part 1 of 2)}
\label{tab:ml_3stage_part1}
\setlength{\tabcolsep}{4pt}
\renewcommand{\arraystretch}{0.95}
\begin{adjustbox}{max width=\linewidth}
\small
\begin{tabular}{lcccccc}
\toprule
 & \multicolumn{3}{c}{\textbf{Column (A)}} & \multicolumn{3}{c}{\textbf{Column (B)}} \\
\cmidrule(lr){2-4} \cmidrule(lr){5-7}
 & \textbf{Small} & \textbf{Large} & \textbf{All} & \textbf{Small} & \textbf{Large} & \textbf{All} \\
\midrule
\multicolumn{7}{l}{\textit{3\textsuperscript{rd} Stage Equation — Outcomes: (A) Log profit margin \quad (B) Log labour productivity}} \\[2pt]
\midrule
\rowcolor[gray]{0.95}
\# of AI devices           & $-0.033$ & $0.006$  & $0.014$       & $0.103^{***}$  & $-0.015$ & $0.153^{***}$ \\
\rowcolor[gray]{0.95}
                           & $(0.093)$ & $(0.014)$ & $(0.014)$     & $(0.036)$      & $(0.111)$ & $(0.029)$ \\
Log $K$                    & $0.006$  & $0.004$  & $0.020^{***}$ & $-0.033^{*}$   & $-0.003$ & $-0.017$ \\
                           & $(0.008)$ & $(0.010)$ & $(0.006)$     & $(0.019)$      & $(0.015)$ & $(0.014)$ \\
\rowcolor[gray]{0.95}
Missing $K$ dummy          & $1.170$  & $0.237$  & $2.432^{***}$ & $-7.873^{***}$ & --       & $-6.341^{***}$ \\
\rowcolor[gray]{0.95}
                           & $(0.888)$ & $(1.091)$ & $(0.662)$     & $(2.461)$      &          & $(2.096)$ \\
Log $L$                    & $0.012$  & $0.058^{***}$ & $0.003$  & $0.018$        & $0.268^{***}$ & $0.073^{***}$ \\
                           & $(0.023)$ & $(0.019)$ & $(0.010)$     & $(0.033)$      & $(0.036)$ & $(0.019)$ \\
\rowcolor[gray]{0.95}
Log Patenting              & $0.010^{**}$ & $0.005$ & $0.007^{**}$ & $0.029^{***}$  & $-0.007$ & $0.025^{***}$ \\
\rowcolor[gray]{0.95}
                           & $(0.005)$ & $(0.005)$ & $(0.003)$     & $(0.007)$      & $(0.007)$ & $(0.006)$ \\
Start-up                   & $-0.251$ & --       & $-0.235$      & $-0.999^{**}$  & --       & $-0.895^{***}$ \\
                           & $(0.158)$ &          & $(0.156)$     & $(0.246)$      &          & $(0.242)$ \\
\rowcolor[gray]{0.95}
New entrant                & --       & --       & --            & --             & --       & -- \\
\rowcolor[gray]{0.95}
                           &          &          &               &                &          &    \\
Country FE ($p$-val.)      & $0.0000$ & $0.0000$ & $0.0000$      & $0.0000$       & $0.0000$ & $0.0000$ \\
\rowcolor[gray]{0.95}
Industry FE ($p$-val.)     & $0.0000$ & $0.0000$ & $0.0000$      & $0.0000$       & $0.0000$ & $0.0000$ \\
Year FE ($p$-val.)         & $0.0780$ & $0.1769$ & $0.0152$      & $0.1344$       & $0.7568$ & $0.0711$ \\
\midrule
\multicolumn{7}{l}{\textit{2\textsuperscript{nd} Stage Equation (Tobit) — Outcome: \# of AI devices introduced (cumulated)}} \\[2pt]
\midrule
\rowcolor[gray]{0.95}
Collaborations in AI       & $0.881^{**}$  & $1.531^{***}$ & $0.755^{**}$  & $0.875^{**}$   & $1.321^{***}$ & $0.758^{**}$ \\
\rowcolor[gray]{0.95}
~~research (average)       & $(0.397)$ & $(0.393)$ & $(0.299)$     & $(0.398)$      & $(0.455)$ & $(0.299)$ \\
Log $K$                    & $0.006$   & $-0.298^{**}$ & $-0.105$  & $0.003$        & $-0.288^{**}$ & $-0.116$ \\
                           & $(0.095)$ & $(0.116)$ & $(0.074)$     & $(0.094)$      & $(0.117)$ & $(0.075)$ \\
\rowcolor[gray]{0.95}
Missing $K$ dummy          & $-1.014$  & $-28.983^{**}$ & $-12.482$ & $-2.043$       & $-28.115^{**}$ & $-13.473^{*}$ \\
\rowcolor[gray]{0.95}
                           & $(9.843)$ & $(12.565)$ & $(7.753)$    & $(9.876)$      & $(12.618)$ & $(7.769)$ \\
Log $L$                    & $0.679^{***}$ & $-0.644^{*}$ & $0.282^{**}$ & $0.696^{***}$  & $-0.473$ & $0.295^{**}$ \\
                           & $(0.172)$ & $(0.346)$ & $(0.118)$     & $(0.173)$      & $(0.437)$ & $(0.119)$ \\
\rowcolor[gray]{0.95}
Log AI P.A.F.              & $0.191^{**}$  & $0.558^{***}$ & $0.149^{**}$ & $0.195^{**}$   & $0.520^{***}$ & $0.152^{***}$ \\
\rowcolor[gray]{0.95}
                           & $(0.082)$ & $(0.151)$ & $(0.066)$     & $(0.083)$      & $(0.162)$ & $(0.066)$ \\
Log Patenting              & $0.203^{***}$ & $-0.694^{***}$ & $0.105^{***}$ & $0.206^{***}$  & $-0.685^{***}$ & $0.110^{***}$ \\
                           & $(0.051)$ & $(0.162)$ & $(0.038)$     & $(0.052)$      & $(0.162)$ & $(0.039)$ \\
\rowcolor[gray]{0.95}
Start-up                   & $1.866^{**}$  & --       & $1.477^{**}$  & $1.891^{**}$   & --       & $1.507^{**}$ \\
\rowcolor[gray]{0.95}
                           & $(0.776)$ &          & $(0.740)$     & $(0.783)$      &          & $(0.744)$ \\
New entrant                & $0.555$   & $-7.030^{*}$ & $0.288$   & $0.543$        & $-7.030^{*}$ & $0.284$ \\
                           & $(0.690)$ & $(4.042)$ & $(0.674)$     & $(0.691)$      & $(4.042)$ & $(0.676)$ \\
\rowcolor[gray]{0.95}
Country FE ($p$-val.)      & $0.0848$  & $0.4346$ & $0.4354$      & $0.0843$       & $0.4478$ & $0.3810$ \\
Industry FE ($p$-val.)     & $0.0001$  & $0.0005$ & $0.0000$      & $0.0001$       & $0.0004$ & $0.0000$ \\
\rowcolor[gray]{0.95}
Year FE ($p$-val.)         & $0.0193$  & $0.0432$ & $0.0000$      & $0.0194$       & $0.0265$ & $0.0000$ \\
\bottomrule
\end{tabular}
\end{adjustbox}
\end{table}

\begin{table}[htbp]
\centering
\addtocounter{table}{-1} 
\caption{3-stage model, simultaneous estimation by ML (Part 2 of 2)}
\label{tab:ml_3stage_part2}
\setlength{\tabcolsep}{4pt}
\renewcommand{\arraystretch}{0.95}
\begin{adjustbox}{max width=\linewidth}
\small
\begin{tabular}{lcccccc}
\toprule
 & \multicolumn{3}{c}{\textbf{Column (A)}} & \multicolumn{3}{c}{\textbf{Column (B)}} \\
\cmidrule(lr){2-4} \cmidrule(lr){5-7}
 & \textbf{Small} & \textbf{Large} & \textbf{All} & \textbf{Small} & \textbf{Large} & \textbf{All} \\
\midrule
\multicolumn{7}{l}{\textit{1\textsuperscript{st} Stage Equation (linear) — Outcome: Collaborations in AI research (average)}} \\[2pt]
\midrule
\rowcolor[gray]{0.95}
Log $K$                    & $0.003$   & $0.017$  & $0.005$       & $0.003$        & $0.017$  & $0.005$ \\
\rowcolor[gray]{0.95}
                           & $(0.005)$ & $(0.016)$ & $(0.006)$     & $(0.005)$      & $(0.016)$ & $(0.006)$ \\
Missing $K$ dummy          & $0.321$   & $1.467$  & $0.420$       & $0.312$        & $1.501$  & $0.420$ \\
                           & $(0.527)$ & $(1.698)$ & $(0.605)$     & $(0.527)$      & $(1.696)$ & $(0.605)$ \\
\rowcolor[gray]{0.95}
Log $L$                    & $0.032^{***}$ & $-0.022$ & $0.018^{**}$  & $0.032^{***}$  & $-0.020$ & $0.018^{**}$ \\
\rowcolor[gray]{0.95}
                           & $(0.009)$ & $(0.056)$ & $(0.008)$     & $(0.009)$      & $(0.055)$ & $(0.008)$ \\
Log AI P.A.F.              & $-0.007$  & $0.012$  & $-0.0005$     & $-0.007$       & $0.011$  & $-0.0005$ \\
                           & $(0.007)$ & $(0.010)$ & $(0.006)$     & $(0.007)$      & $(0.011)$ & $(0.006)$ \\
\rowcolor[gray]{0.95}
Log Patenting              & $0.004^{**}$  & $0.019^{***}$ & $0.008^{***}$ & $0.004^{**}$   & $0.020^{***}$ & $0.008^{***}$ \\
\rowcolor[gray]{0.95}
                           & $(0.002)$ & $(0.007)$ & $(0.002)$     & $(0.002)$      & $(0.007)$ & $(0.002)$ \\
Start-up                   & $0.035$   & --       & $0.023$       & $0.036$        & --       & $0.023$ \\
                           & $(0.038)$ &          & $(0.040)$     & $(0.038)$      &          & $(0.040)$ \\
\rowcolor[gray]{0.95}
New entrant                & $-0.026$  & $0.860$  & $-0.015$      & $-0.026$       & $0.888$  & $-0.015$ \\
\rowcolor[gray]{0.95}
                           & $(0.025)$ & $(0.703)$ & $(0.025)$     & $(0.025)$      & $(0.703)$ & $(0.025)$ \\
Distance to AI hub (km)    & $-0.00002$ & $-0.00004$ & $-0.00001$   & $-0.00002$     & $-0.0001$ & $-0.00001$ \\
                           & $(0.00002)$ & $(0.0001)$ & $(0.00002)$  & $(0.00002)$    & $(0.0001)$ & $(0.00002)$ \\
\rowcolor[gray]{0.95}
Log AI publications        & $2.889^{***}$ & $1.446^{***}$ & $1.831^{***}$ & $2.889^{***}$  & $1.430^{***}$ & $1.829^{***}$ \\
\rowcolor[gray]{0.95}
                           & $(0.270)$ & $(0.101)$ & $(0.096)$     & $(0.270)$      & $(0.098)$ & $(0.095)$ \\
Country FE ($p$-val.)      & $0.0020$  & $0.4990$ & $0.0000$      & $0.0017$       & $0.4639$ & $0.0000$ \\
\rowcolor[gray]{0.95}
Industry FE ($p$-val.)     & $0.2398$  & $0.0000$ & $0.4210$      & $0.2410$       & $0.0000$ & $0.4207$ \\
Year FE ($p$-val.)         & $0.0123$  & $0.2722$ & $0.0010$      & $0.0126$       & $0.2513$ & $0.0010$ \\
\midrule
\multicolumn{7}{l}{\textit{Ancillary Model Parameters}} \\[2pt]
\midrule
\rowcolor[gray]{0.95}
$\log\sigma_1$             & $-0.186$  & $0.611^{***}$ & $0.122$       & $-0.186$       & $0.611^{***}$ & $0.122$ \\
\rowcolor[gray]{0.95}
                           & $(0.125)$ & $(0.189)$ & $(0.111)$     & $(0.125)$      & $(0.189)$ & $(0.111)$ \\
$\log\sigma_2$             & $1.868^{***}$ & $2.088^{***}$ & $1.967^{***}$ & $1.868^{***}$  & $2.077^{***}$ & $1.967^{***}$ \\
                           & $(0.200)$ & $(0.152)$ & $(0.145)$     & $(0.200)$      & $(0.150)$ & $(0.145)$ \\
\rowcolor[gray]{0.95}
$\log\sigma_3$             & $0.058^{***}$ & $-0.163^{***}$ & $-0.0015$   & $0.378^{***}$  & $-0.032$ & $0.330^{***}$ \\
\rowcolor[gray]{0.95}
                           & $(0.020)$ & $(0.035)$ & $(0.018)$     & $(0.040)$      & $(0.074)$ & $(0.035)$ \\
$\operatorname{atanh}\rho_{12}$ & $-0.100^{**}$ & $-0.379^{***}$ & $-0.084^{*}$  & $-0.099^{**}$  & $-0.339^{***}$ & $-0.085^{*}$ \\
                           & $(0.048)$ & $(0.123)$ & $(0.047)$     & $(0.049)$      & $(0.131)$ & $(0.047)$ \\
\rowcolor[gray]{0.95}
$\operatorname{atanh}\rho_{13}$ & $0.015$   & $-0.025$ & $-0.004$      & $0.029^{**}$   & $0.040$  & $0.012$ \\
\rowcolor[gray]{0.95}
                           & $(0.013)$ & $(0.022)$ & $(0.011)$     & $(0.012)$      & $(0.031)$ & $(0.010)$ \\
$\operatorname{atanh}\rho_{23}$ & $-0.061$  & $-0.013$ & $-0.061$      & $-0.135^{***}$ & $0.314$  & $-0.141^{***}$ \\
                           & $(0.075)$ & $(0.073)$ & $(0.056)$     & $(0.039)$      & $(0.258)$ & $(0.032)$ \\
\midrule
\rowcolor[gray]{0.95}
Observations               & $7{,}515$ & $1{,}807$ & $9{,}322$     & $7{,}515$      & $1{,}807$ & $9{,}322$ \\
Firms                      & $3{,}980$ & $262$    & $4{,}186$     & $3{,}980$      & $262$    & $4{,}186$ \\
\rowcolor[gray]{0.95}
Log-likelihood             & $-13879.29$ & $-5795.40$ & $-21273.90$ & $-14585.64$    & $-4982.90$ & $-21235.58$ \\
\bottomrule
\end{tabular}
\end{adjustbox}
\begin{minipage}{\linewidth}
\smallskip\footnotesize
\textit{Notes:} Significance levels: $^{***}$ 1\%, $^{**}$ 5\%, $^{*}$ 10\%. Robust standard errors are reported in parentheses. FE lines display the $p$-value from an $F$-test of joint significance ($H_0$: FE coefficients are not significant). The parameters $\sigma_1$, $\sigma_2$, and $\sigma_3$ represent the standard deviations of the error terms for equations 1--3, respectively, while $\rho_{12}$, $\rho_{13}$, and $\rho_{23}$ capture the cross-equation error correlations. The ML framework yields explicit estimates of $\operatorname{atanh}\rho_{ij}$ and $\log\sigma_i$. LR tests systematically reject the null hypothesis of all-zero coefficients across specifications. In the ``Large'' subsample of Column (B), the ``Missing $K$ dummy'' is identically zero when Value Added is the outcome, hence its coefficient is omitted. The number of observations is slightly larger than in the sequential estimations (i) because the endogenous regressor in the third stage is not predicted but observed and (ii) because the CMP procedure uses all available observations per equation.
\end{minipage}
\end{table}

Table~\ref{tab:ml_3stage_part1} reports the results of the FIML estimator implemented via the CMP routine. All three equations are estimated jointly, allowing for unrestricted correlation across the error terms of the collaboration, device introduction, and firm performance equations. This specification serves as a robustness check on the sequential IV~Tobit results reported in Section~\ref{TobitSeq}.

The second-stage results confirm the IV~Tobit findings. Collaborations in AI research retain a positive and significant effect on the cumulative number of AI devices introduced across all sub-samples and both outcome columns. The error correlation parameter $\operatorname{atanh}\hat{\rho}_{12}$ is negative and significant across all subsamples, confirming endogeneity of AI collaborations and validating the instrumentation strategy throughout.

The third-stage results for labour productivity closely mirror those of the sequential estimator. The number of AI devices brought to the market has a positive and significant effect among small firms and in the full sample in Column~(B), whereas the large-firm coefficient is negative and non-significant. For profit margins in Column~(A), the device count is not significant in any subsample under joint estimation, consistent with the weaker and less stable pattern already observed in the sequential results. The FIML estimates confirm that the productivity and adoption results obtained under sequential IV~Tobit estimation are robust to changes in the estimator and estimation approaches.

\section{Conclusion and future work}
\label{sec:conclusion}

This paper examined the full innovation chain linking AI capability-building to market outcomes in the US medical device industry. Using a novel dataset that connects FDA clearance records, patent data, scientific publications, and firm financials, we estimated a three-stage recursive model that traces the process from external collaboration through AI device introduction to firm performance. The model allows us to identify -- in the whole sample and for small firms -- a full linkage from collaborations in AI research to the introduction of AI devices to labour productivity.

Looking at the estimates in more details, three findings stand out. First, external collaboration is a significant driver of AI device introduction across all firm sizes and estimators. The effect is larger for small firms. This may be explained with external knowledge ties that substitute for limited internal R\&D capacity. Second, AI device adoption raises labour productivity robustly. The productivity premium is significant for small firms and the full sample across all specifications, and holds under both sequential and joint ML estimation. Third, profit margin effects are present but less robust. They emerge for the full sample and for large firms under the count-based Tobit specification, but do not survive joint estimation uniformly. This pattern is consistent with competitive entry eroding pricing power as AI devices diffuse through the sector, while productivity gains persist because they are embedded in the firm's production process rather than extracted through product pricing.

Among the controls included in the models, the role of medtech start-ups is worth mentioning. They have a higher probability of introducing an AI innovation but the estimations of the third stage indicate a systematic negative productivity correlation and a mostly not significant negative profit margin. The estimation is consistent with early-stage scaling dynamics for small young firms.

These results carry implications beyond the immediate estimates. First, the size-dependence of the collaboration effect, 4.7 percentage points for small firms against 2.8 for large firms in the LPM specification (Table 2), suggests that policies aimed at accelerating AI diffusion in medtech — matching grants, collaborative R\&D tax credits, or subsidised access to shared clinical data infrastructure — will have the largest marginal impact if targeted at smaller, resource-constrained firms rather than incumbents that already possess the internal capacity to substitute for external ties. The partner-type decomposition (Table S1) sharpens this point: because company and healthcare partnerships carry larger marginal effects than academic ties despite being the rarer forms of collaboration, instruments that specifically de-risk industry-to-industry or industry-to-clinic agreements, and not only university technology-transfer channels, may be more cost-effective per device brought to market than research-collaboration subsidies alone.

The estimation on the complementarity between firm size and productivity extraction per device (Table 6) provides some information on the aggregate productivity payoff from AI diffusion in medtech: large firms are those that seem better in extracting productivity from multiple AI-enhanced devices (though this result should be taken with care as it does not survive the joint ML estimator). If deep AI device portfolios remain concentrated among small firms that are effective at entry but extract comparatively less productivity per device, aggregate gains will be smaller than if the same devices would be introduced by large companies. This result indicates that we could expect to see a process of technology transfer from small innovators to large developers — through licensing, acquisition, or partnership — similar to what has characterized the biotech industry.

Several limitations qualify how these results should be read. First, the sample is heavily US-weighted (72\% of matched firms), and the FDA list captures devices cleared almost exclusively through the 510(k) pathway (94\%). Second, the collaboration and publication measures are constructed for firms worldwide, but the innovation-output measure is anchored to a single, US-specific regulatory route, so the results speak most directly to firms operating under, or seeking clearance through, that architecture, and less directly to firms whose primary AI device pathway runs through the EU's MDR/IVDR/AI Act framework. Third, financial-performance data are also considerably sparser than the collaboration and device data: the third-stage estimation samples range from 2,976 to 3,694 firm-year observations, against 8,774 in the second stage, because Orbis coverage of profit margin and value added is incomplete. If the firms with usable financial data differ systematically (in transparency, listing status, or maturity) from those without, the productivity and margin estimates carry a residual selection margin beyond the endogeneity already addressed by instrumentation. The unbalanced structure of the panel — a large share of firms appear only once due to lack of coverage by Orbis — rules out individual fixed effects. Fourth, the collaboration measure itself is built from co-authored AI publications, and therefore captures only collaboration that produces a joint scientific output. Collaboration channels that firms have a commercial incentive to keep undisclosed such as licensing agreements, data-sharing contracts, or consulting arrangements protected by non-disclosure terms are not observed. Since company and clinical partnerships are plausibly the channels most likely to be governed by such agreements, the partner-type decomposition should be read as a lower bound on the true intensity of non-academic collaboration, not a full account of it.  Finally, the full-sample specification does not clear the overidentification test at the 5\% level (Hansen J = 5.854, p = 0.016), even though it does so comfortably within each firm-size subsample; this should temper confidence in the pooled full-sample coefficients relative to the size-stratified estimates, which remain the more conservative reading of the results. 


\clearpage

\printbibliography

@article{SeWo2010,
  author    = {Semykina, A. and J.K. Wooldridge},
  title     = {Estimating panel data models in the presence of endogeneity and selection},
  journal   = {Journal of Econometrics},
  year      = {2010},
  volume    = {157},
  pages     = {375--380}
}

@article{New1987,
  author    = {Newey, W. K.},
  title     = {Efficient estimation of limited dependent variable models with endogenous explanatory variables},
  journal   = {Journal of Econometrics},
  year      = {1987},
  volume    = {36},
  pages     = {231--250}
}

@article{Agrawal2019,
  title={The Economics of Artificial Intelligence: An Agenda},
  author={Ajay Agrawal and Joshua S. Gans and Avi Goldfarb},
  year={2019},
  journal={RePEc: Research Papers in Economics},
  doi={10.7208/chicago/9780226613475.001.0001}
}

@article{Bianchini2022,
  title={Artificial intelligence in science: An emerging general method of invention},
  author={Bianchini, Stefano and Müller, Moritz and Pelletier, Pierre},
  year={2022},
  journal={Research Policy},
  doi={10.1016/j.respol.2022.104604}
}

@article{Cockburn2018,
  title={The Impact of Artificial Intelligence on Innovation},
  author={Cockburn, Iain and Henderson, Rebecca and Stern, Scott},
  year={2018},
  doi={10.3386/w24449}
}

@misc{HTR2025,
  author    = {{Healthcare Technology Report}},
  title     = {The Top 25 Healthcare {AI} Companies of 2025},
  year      = {2025},
  howpublished = {\url{https://thehealthcaretechnologyreport.com/the-top-25-healthcare-ai-companies-of-2025/}}
}

@misc{Landi2025,
  author    = {Landi, H.},
  title     = {Healthcare {AI} rakes in nearly \${4B} in {VC} funding, buoying the digital health market in 2025},
  journal   = {FierceHealthcare},
  year      = {2025},
  howpublished = {\url{https://www.fiercehealthcare.com/health-tech/healthcare-ai-rakes-nearly-4b-vc-funding-buoying-digital-health-market-2025}}
}

@misc{Precedence2026, % web
  author       = {{Precedence Research}},
  title        = {Artificial Intelligence in Healthcare Market Size to Hit {USD} 744.34 {Bn} by 2035},
  year         = {2026},
  howpublished = {\url{https://www.precedenceresearch.com/artificial-intelligence-in-healthcare-market}},
  note         = {Accessed on 31.07.2026}
}

@misc{Tempus2025,
  author       = {{Tempus AI, Inc.}},
  title        = {Tempus announces the acquisition of {Paige}},
  year         = {2025},
  howpublished = {\url{https://investors.tempus.com/news-releases/news-release-details/tempus-announces-acquisition-paige}},
  note         = {Accessed: June 2025}
}

@misc{Tempus2025b,
  author       = {{Tempus AI, Inc.}},
  title        = {Tempus completes acquisition of {Ambry} {Genetics}},
  year         = {2025},
  howpublished = {\url{https://investors.tempus.com/news-releases/news-release-details/tempus-completes-acquisition-ambry-genetics}},
  note         = {Accessed: July 2025}
}

@article{Brynjolfsson2025,
  author  = {Brynjolfsson, Erik and Li, Danielle and Raymond, Lindsey},
  title   = {Generative {AI} at Work},
  journal = {Quarterly Journal of Economics},
  year    = {2025},
  note    = {qjae044}
}

@article{Comunale2024, % Working paper
  title={The Economic Impacts and the Regulation of AI: A Review of the Academic Literature and Policy Actions},
  author={Mariarosaria Comunale},
  year={2024},
  journal={IMF Working Paper},
  doi={10.5089/9798400268588.001}
}

@article{AcemogluRestrepo2020,
  title={Artificial Intelligence, Automation, and Work},
  author={Acemoglu, Daron and Restrepo, Pascual},
  year={2019},
  journal={The Economics of Artificial Intelligence},
  doi={10.7208/chicago/9780226613475.003.0008}
}

@article{CirilloMina2023,
  author  = {Cirillo, Valeria and Fanti, Lucrezia and Mina, Andrea and Ricci, Andrea},
  title   = {New Digital Technologies and Firm Performance in the {Italian} Economy},
  journal = {Industry and Innovation},
  year    = {2023},
  volume  = {30},
  number  = {1},
  pages   = {159--188}
}

@article{Griffith2006,
  author  = {Griffith, Rachel and Huergo, Elena and Mairesse, Jacques and Peters, Bettina},
  title   = {Innovation and Productivity across Four {European} Countries},
  journal = {Oxford Review of Economic Policy},
  year    = {2006},
  volume  = {22},
  number  = {4},
  pages   = {483--498},
  doi = {https://doi.org/10.1093/oxrep/grj028}
}

@article{HallLottiMairesse2009,
  author  = {Hall, B.H. and Lotti, Francesca and Mairesse, Jacques},
  title   = {Innovation and Productivity in {SMEs}: Empirical Evidence for {Italy}},
  journal = {Small Business Economics},
  year    = {2009},
  volume  = {33},
  number  = {1},
  pages   = {13--33}
}

@article{Crepon1998,
  title={Research, Innovation And Productivity: An Econometric Analysis At The Firm Level},
  author={Crepon, Bruno and Duguet, Emmanuel and Mairesse, Jacques},
  year={1998},
  journal={Economics of Innovation and New Technology},
  doi={10.1080/10438599800000031}
}

@article{Davenport2018,
  author  = {Davenport, T. H. and Ronanki, R.},
  title   = {Artificial intelligence for the real world},
  journal = {Harvard Business Review},
  year    = {2018},
  volume  = {96},
  number  = {1},
  pages   = {108--116}
}

@article{Babina2024,
  author    = {Babina, T. and Fedyk, A. and He, A. and Hodson, J.},
  title     = {Artificial intelligence, firm growth, and product innovation},
  journal   = {Journal of Financial Economics},
  year      = {2024},
  volume    = {151},
  pages     = {103745},
  doi       = {10.1016/j.jfineco.2023.103745}
}

@article{Cohen1990,
  author    = {Cohen, Wesley M. and Levinthal, Daniel A.},
  title     = {{Absorptive Capacity}: A New Perspective on Learning and Innovation},
  journal   = {Administrative Science Quarterly},
  year      = {1990},
  volume    = {35},
  number    = {1},
  pages     = {128--152},
  doi       = {10.2307/2393553}
}

@article{Bianchini2025AIAS,
  title={Scientific discovery in the age of AI and supercomputing},
  author={Stefano Bianchini and Aldo Geuna and Fazliddin Shermatov},
  year={2026},
  journal={Scientific Reports},
  doi={10.1038/s41598-026-63438-7}
}

@article{Roodman2011,
  title={Fitting Fully Observed Recursive Mixed-process Models with cmp},
  author={Roodman, David},
  year={2011},
  journal={The Stata Journal: Promoting communications on statistics and Stata},
  doi={10.1177/1536867x1101100202}
}

@article{Pairolero2025,
  title={The artificial intelligence patent dataset (AIPD) 2023 update},
  author={Pairolero, Nicholas A. and Giczy, Alexander V. and Torres, Gerard and Islam Erana, Tisa and Finlayson, Mark A. and Toole, Andrew A.},
  year={2025},
  journal={The Journal of Technology Transfer},
  doi={10.1007/s10961-025-10189-8}
}

@article{Roppelt2024,
  title={Artificial intelligence in healthcare institutions: A systematic literature review on influencing factors},
  author={Roppelt, Julia Stefanie and Kanbach, Dominik K. and Kraus, Sascha},
  year={2024},
  journal={Technology in Society},
  doi={10.1016/j.techsoc.2023.102443}
}

@article{Singh2020,
  title={Current Challenges and Barriers to Real-World Artificial Intelligence Adoption for the Healthcare System, Provider, and the Patient},
  author={Singh, Rishi P. and Hom, Grant L. and Abramoff, Michael D. and Campbell, J. Peter and Chiang, Michael F. and undefined},
  year={2020},
  journal={Translational Vision Science amp; Technology},
  doi={10.1167/tvst.9.2.45}
}

@article{Pianykh2020,
  author    = {Pianykh, O. S. and Langs, G. and Dewey, M. and Enzmann, D. R. and Herold, C. J. and Schoenberg, S. O. and Brink, J. A.},
  title     = {Continuous learning {AI} in radiology: implementation principles and early applications},
  journal   = {Radiology},
  year      = {2020},
  volume    = {297},
  number    = {1},
  pages     = {6--14},
  doi       = {10.1148/radiol.2020200038}
}

@article{Noorbakhsh2019,
  author    = {Noorbakhsh-Sabet, N. and Zand, R. and Zhang, Y. and Abedi, V.},
  title     = {Artificial intelligence transforms the future of health care},
  journal   = {The American Journal of Medicine},
  year      = {2019},
  volume    = {132},
  number    = {7},
  pages     = {795--801},
  doi       = {10.1016/j.amjmed.2019.01.017}
}

@article{Watson2020,
  title={Overcoming barriers to the adoption and implementation of predictive modeling and machine learning in clinical care: what can we learn from US academic medical centers?},
  author={Watson, Joshua and Hutyra, Carolyn A and Clancy, Shayna M and Chandiramani, Anisha and Bedoya, Armando and Ilangovan, Kumar and Nderitu, Nancy and Poon, Eric G},
  year={2020},
  journal={JAMIA Open},
  doi={10.1093/jamiaopen/ooz046}
}

@article{Apell2021,
  title={Artificial intelligence (AI) healthcare technology innovations: the current state and challenges from a life science industry perspective},
  author={Apell, Petra and Eriksson, Henrik},
  year={2021},
  journal={Technology Analysis \&; Strategic Management},
  doi={10.1080/09537325.2021.1971188}
}

@misc{AIAct2024, % BE BACK
  author    = {{European Parliament and Council of the European Union}},
  title     = {Regulation ({EU}) 2024/1689 of the {European Parliament} and of the {Council} of 13 {June} 2024 laying down harmonised rules on artificial intelligence (Artificial {Intelligence} {Act})},
  year      = {2024},
  journal   = {Official Journal of the European Union},
  note      = {OJ L, 2024/1689}
}

@misc{EUCouncil2024MDR, % BE BACK
  author    = {{Council of the European Union}},
  title     = {{AI} act: {Council} gives final green light to the first worldwide rules on {AI}},
  year      = {2024},
  howpublished = {\url{https://www.consilium.europa.eu/en/press/press-releases/2024/05/21/ai-act-council-gives-final-green-light-to-the-first-worldwide-rules-on-ai/}}
}

@article{Newhouse1992,
  author    = {Newhouse, J. P.},
  title     = {Medical care costs: how much welfare loss?},
  journal   = {Journal of Economic Perspectives},
  year      = {1992},
  volume    = {6},
  number    = {3},
  pages     = {3--21}
}

@book{Cutler2004,
  author    = {Cutler, D. M.},
  title     = {Your Money or Your Life: Strong Medicine for {America}'s Health Care System},
  publisher = {Oxford University Press},
  year      = {2004}
}

@article{Nordhaus1969, 
  author    = {Nordhaus, W. D.},
  title     = {An economic theory of technological change},
  journal   = {The American Economic Review},
  year      = {1969},
  volume    = {59},
  number    = {2},
  pages     = {18--28}
}

@article{AccemogluLinn2004,
  title={Market Size in Innovation: Theory and Evidence from the Pharmaceutical Industry},
  author={Acemoglu, D. and Linn, J.},
  year={2004},
  journal={The Quarterly Journal of Economics},
  doi={10.1162/0033553041502144}
}

@article{Finkelstein2004,
  title={Static and Dynamic Effects of Health Policy: Evidence from the Vaccine Industry},
  author={Finkelstein, A.},
  year={2004},
  journal={The Quarterly Journal of Economics},
  doi={10.1162/0033553041382166}
}

@article{WardDranove1995,
  author    = {Ward, M. R. and Dranove, D.},
  title     = {The vertical chain of research and development in the pharmaceutical industry},
  journal   = {Economic Inquiry},
  year      = {1995},
  volume    = {33},
  number    = {1},
  pages     = {70--87}
}

@article{AIdiffusion2024,
  author    = {Agrawal, A. and Gans, J. and Goldfarb, A.},
  title     = {Artificial intelligence adoption and system-wide change},
  journal   = {Journal of Economics \& Management Strategy},
  year      = {2024},
  volume    = {33},
  number    = {2},
  pages     = {327--337},
  doi       = {10.1111/jems.12521}
}

@article{Crafts2021,
  author  = {Crafts, Nicholas},
  title   = {Artificial intelligence as a general-purpose technology: an historical perspective},
  journal = {Oxford Review of Economic Policy},
  year    = {2021},
  volume  = {37},
  number  = {3},
  pages   = {521--536},
  publisher = {Oxford University Press and Oxford Review of Economic Policy Limited}
}

@techreport{Eisfeldt2023,
  author      = {Eisfeldt, A. L. and Schubert, G. and Zhang, M. B.},
  title       = {Generative {AI} and Firm Values},
  institution = {National Bureau of Economic Research},
  year        = {2023},
  type        = {Working Paper},
  number      = {31222},
  doi         = {10.3386/w31222}
}

@article{Arora2018,
  author    = {Arora, A. and Belenzon, S. and Patacconi, A.},
  title     = {The decline of science in corporate {R\&D}},
  journal   = {Strategic Management Journal},
  year      = {2018},
  volume    = {39},
  number    = {1},
  pages     = {3--32}
}

@article{Cohen2002,
  author    = {Cohen, W. M. and Nelson, R. R. and Walsh, J. P.},
  title     = {Links and impacts: the influence of public research on industrial {R\&D}},
  journal   = {Management Science},
  year      = {2002},
  volume    = {48},
  number    = {1},
  pages     = {1--23}
}

@article{Arvanitis2008,
  author    = {Arvanitis, S. and Kubli, U. and Woerter, M.},
  title     = {University-industry knowledge and technology transfer in {Switzerland}: what university scientists think about co-operation with private enterprises},
  journal   = {Research Policy},
  year      = {2008},
  volume    = {37},
  number    = {10},
  pages     = {1865--1883}
}

@article{Caloghirou2021,
  author    = {Caloghirou, Y. and Giotopoulos, I. and Kontolaimou, A. and Korra, E. and Tsakanikas, A.},
  title     = {Industry-university knowledge flows and product innovation: how do knowledge stocks and crisis matter?},
  journal   = {Research Policy},
  year      = {2021},
  volume    = {50},
  number    = {3},
  pages     = {104195}
}

@article{Tether2008,
  author    = {Tether, B. S. and Tajar, A.},
  title     = {Beyond industry-university links: sourcing knowledge for innovation from consultants, private research organisations and the public science-base},
  journal   = {Research Policy},
  year      = {2008},
  volume    = {37},
  number    = {6--7},
  pages     = {1079--1095}
}

@article{GonzalezPernia2015,
  title={STI–DUI learning modes, firm–university collaboration and innovation},
  author={González-Pernía, José L. and Parrilli, Mario Davide and Peña-Legazkue, Iñaki},
  year={2014},
  journal={The Journal of Technology Transfer},
  doi={10.1007/s10961-014-9352-0}
}

@article{Perkmann2007,
  author    = {Perkmann, M. and Walsh, K.},
  title     = {University-industry relationships and open innovation: towards a research agenda},
  journal   = {International Journal of Management Reviews},
  year      = {2007},
  volume    = {9},
  number    = {4},
  pages     = {259--280}
}

@article{Perkmann2013,
  title={Academic engagement and commercialisation: A review of the literature on university–industry relations},
  author={Perkmann, Markus and Tartari, Valentina and McKelvey, Maureen and Autio, Erkko and Broström, Anders and D’Este, Pablo and Fini, Riccardo and Geuna, Aldo and Grimaldi, Rosa and Hughes, Alan and Krabel, Stefan and Kitson, Michael and Llerena, Patrick and Lissoni, Franceso and Salter, Ammon and Sobrero, Maurizio},
  year={2013},
  journal={Research Policy},
  doi={10.1016/j.respol.2012.09.007}
}

@article{YuLee2017,
  author    = {Yu, G. J. and Lee, J.},
  title     = {When should a firm collaborate with research organizations for innovation performance? The moderating role of innovation orientation, size, and age},
  journal   = {Journal of Technology Transfer},
  year      = {2017},
  volume    = {42},
  number    = {6},
  pages     = {1451--1465}
}

@article{Szucs2018,
  author    = {Sz{\"u}cs, F.},
  title     = {Research subsidies, industry-university cooperation and innovation},
  journal   = {Research Policy},
  year      = {2018},
  volume    = {47},
  number    = {7},
  pages     = {1256--1266}
}

@article{VegaJurado2017,
  author    = {Vega-Jurado, J. and Kask, S. and Manjarr{\'e}s-Henriquez, L.},
  title     = {University industry links and product innovation: cooperate or contract},
  journal   = {Journal of Technology Management \& Innovation},
  year      = {2017},
  volume    = {12},
  number    = {3},
  pages     = {1--8}
}

@article{GarciaVega2020,
  author    = {Garc{\'i}a-Vega, M. and Vicente-Chirivella, O.},
  title     = {Do university technology transfers increase firms' innovation?},
  journal   = {European Economic Review},
  year      = {2020},
  volume    = {123},
  pages     = {103388}
}

@article{AnonHigon2016,
  author    = {{Anon Hig{\'o}n}, D. A.},
  title     = {In-house versus external basic research and first-to-market innovations},
  journal   = {Research Policy},
  year      = {2016},
  volume    = {45},
  number    = {4},
  pages     = {816--829}
}

@techreport{BargeGil2019, % working paper
  author      = {Barge-Gil, A. and Vivas-Augier, C.},
  title       = {Does Cooperation with Universities and {KIBS} Matter? Firm-level Evidence from {Spain}},
  year        = {2019},
  institution = {University Library of Munich},
  type        = {MPRA Paper},
  number      = {96949}
}

@article{Hall2011,
  author  = {Hall, B.H.},
  title   = {Innovation and productivity},
  journal = {Nordic Economic Policy Review},
  year    = {2011},
  volume  = {2},
  pages   = {167--204}
}

@article{Fudickar2019,
  author    = {Fudickar, R. and Hottenrott, H.},
  title     = {Public research and the innovation performance of new technology based firms},
  journal   = {Journal of Technology Transfer},
  year      = {2019},
  volume    = {44},
  number    = {2},
  pages     = {326--358}
}

@article{CetteNevoux2022,
  author  = {Cette, Gilbert and Nevoux, Sandra and Py, Loriane},
  title   = {The impact of {ICTs} and digitalization on productivity and labor share: evidence from {French} firms},
  journal = {Economics of Innovation and New Technology},
  year    = {2022},
  volume  = {31},
  number  = {8},
  pages   = {669--692},
  doi     = {10.1080/10438599.2020.1849967}
}

@article{CalvinoFontanelli2026,
  author  = {Calvino, Flavio and Fontanelli, Luca},
  title   = {{AI} users are not all alike: The characteristics of {French} firms buying and developing {AI}},
  journal = {Research Policy},
  year    = {2026},
  volume  = {55},
  number  = {5},
  pages   = {105473},
  doi     = {10.1016/j.respol.2026.105473}
}

@incollection{HallMairesse2010,
  author    = {Hall, B.H. and Mairesse, J. and Mohnen, P.},
  title     = {Measuring the Returns to {R\&D}},
  booktitle = {Handbook of the Economics of Innovation},
  editor    = {Hall, B.H. and Rosenberg, N.},
  publisher = {Elsevier},
  year      = {2010},
  doi       = {10.1016/s0169-7218(10)02008-3}
}

@incollection{MairesseRobin2012,
  author    = {Mairesse, J. and Robin, S.},
  title     = {The Importance of Research for Innovation and Productivity: Comparing Different Estimators of the Innovation Production Function},
  booktitle = {Innovation and Growth: From R\&D Strategies of Innovating Firms to Economy-wide Technological Change},
  editor    = {Andersson, M. and B. Johansson and C. Karlsson and H. Lööf},
  publisher = {Oxford University Press},
  year      = {2012},
  pages     = {368 p.},
  doi       = {https://doi.org/10.1093/acprof:oso/9780199646685.001.0001}
}

@article{LoofMairesseMohnen2017,
  author    = {Lööf, H. and Mairesse, J. and Mohnen, P.},
  title     = {CDM 20 years after},
  journal   = {Economics of Innovation and New Technology},
  year      = {2017},
  volume    = {26},
  number    = {1-2},
  pages     = {1-5},
  doi = {https://doi.org/10.1080/10438599.2016.1202522}
}

@article{NottenMairesseVerspagen2017,
  author    = {Notten, A. and Mairesse, J. and Verspagen, B.},
  title     = {The CDM framework: knowledge recombination from an evolutionary viewpoint},
  journal   = {Economics of Innovation and New Technology},
  year      = {2017},
  volume    = {26},
  number    = {1-2},
  pages     = {21-41},
  doi = {https://doi.org/10.1080/10438599.2016.1202520}
}

@article{MairesseRobin2017,
  author    = {Mairesse, J. and Robin, S.},
  title     = {Assessing measurement errors in the CDM research–innovation–productivity relationships},
  journal   = {Economics of Innovation and New Technology},
  year      = {2017},
  volume    = {26},
  number    = {1-2},
  pages     = {93-107},
  doi = {https://doi.org/10.1080/10438599.2016.1210771}
}

@article{HallSena2017,
  author    = {Hall, B.H. and Sena, V.},
  title     = {Appropriability mechanisms, innovation, and productivity: evidence from the UK},
  journal   = {Economics of Innovation and New Technology},
  year      = {2017},
  volume    = {26},
  number    = {1-2},
  pages     = {42-62},
  doi = {https://doi.org/10.1080/10438599.2016.1202513}
}

@article{vanLeeuwenMohnen2017,
  author    = {van Leeuwen, G. and Mohnen, P.},
  title     = {Revisiting the Porter hypothesis: an empirical analysis of Green innovation for the Netherlands},
  journal   = {Economics of Innovation and New Technology},
  year      = {2017},
  volume    = {26},
  number    = {1-2},
  pages     = {63-77},
  doi = {https://doi.org/10.1080/10438599.2016.1202521}
}

@article{RobinSchubert2013,
  author    = {Robin, S. and Schubert, T.},
  title     = {Cooperation with public research institutions and success in innovation: Evidence from France and Germany},
  journal   = {Research Policy},
  year      = {2013},
  volume    = {42},
  number    = {1},
  pages     = {149-166},
  doi = {https://doi.org/10.1016/j.respol.2012.06.002}
}

@article{Griliches1996,
  author    = {Griliches, Z.},
  title     = {The discovery of the residual: a historical note},
  journal   = {Journal of Economic Literature},
  year      = {1996},
  volume    = {34},
  number    = {3},
  pages     = {1324--1330}
}

@techreport{GraetzMichaels2015,
  author      = {Graetz, G. and Michaels, G.},
  title       = {Robots at Work},
  institution = {Centre for Economic Performance},
  year        = {2015},
  type        = {Discussion Paper},
  number      = {1335}
}

@article{Seamans2018,
  author    = {Seamans, R. and Raj, M.},
  title     = {{AI}, Labor, Productivity and the Need for Firm-Level Data},
  journal   = {NBER Working Paper},
  year      = {2018},
  note      = {No. 24239}
}

@article{Thomason2021,
  author    = {Thomason, J.},
  title     = {Big tech, big data and the new world of digital health},
  journal   = {SSRN},
  year      = {2021},
  doi       = {10.2139/ssrn.3847344}
}

@article{McElheran2024,
  author  = {McElheran, Kristina and Li, J. Frank and Brynjolfsson, Erik
             and Kroff, Zachary and Dinlersoz, Emin and Foster, Lucia
             and Zolas, Nikolas},
  title   = {{AI} adoption in {America}: who, what, and where},
  journal = {Journal of Economics \& Management Strategy},
  year    = {2024},
  volume  = {33},
  number  = {2},
  pages   = {375--415},
  doi     = {10.1111/jems.12576}
}

@misc{Alderucci2020,
  author       = {Alderucci, Dean and Branstetter, Lee and Hovy, Eduard
                  and Runge, Alexander and Zolas, Nikolas},
  title        = {Quantifying the impact of {AI} on productivity and labor
                  demand: evidence from {US} census microdata},
  howpublished = {Allied Social Science Associations Annual Meeting (ASSA)},
  year         = {2020}
}

@article{IgnaVenturini2023,
  author  = {Igna, Ioana and Venturini, Francesco},
  title   = {The determinants of {AI} innovation across {European} firms},
  journal = {Research Policy},
  year    = {2023},
  volume  = {52},
  number  = {2},
  pages   = {104661},
  doi     = {10.1016/j.respol.2022.104661}
}

@unpublished{Alekseeva2020,
  author = {Alekseeva, Liudmila and Gin{\'e}, Mireia and Samila, Sampsa
            and Taska, Bledi},
  title  = {{AI} adoption and firm performance: management versus {IT}},
  note   = {Available at SSRN~3677237},
  year   = {2020}
}

@article{Topol2019,
  author    = {Topol, Eric J.},
  title     = {High-performance medicine: the convergence of human and
               artificial intelligence},
  journal   = {Nature Medicine},
  year      = {2019},
  volume    = {25},
  number    = {1},
  pages     = {44--56},
  doi       = {10.1038/s41591-018-0300-7},
}

@article{Rajpurkar2022,
  author    = {Rajpurkar, Pranav and Chen, Emma and Banerjee, Oishi
               and Topol, Eric J.},
  title     = {{AI} in health and medicine},
  journal   = {Nature Medicine},
  year      = {2022},
  volume    = {28},
  number    = {1},
  pages     = {31--38},
  doi       = {10.1038/s41591-021-01614-0},
}

@article{ObermeierEmanuel2016,
  author    = {Obermeyer, Ziad and Emanuel, Ezekiel J.},
  title     = {Predicting the Future --- {Big Data}, Machine Learning,
               and Clinical Medicine},
  journal   = {New England Journal of Medicine},
  year      = {2016},
  volume    = {375},
  number    = {13},
  pages     = {1216--1219},
  doi       = {10.1056/NEJMp1606181},
}

@article{Benjamens2020,
  author    = {Benjamens, Stan and Dhunnoo, Pranavsingh and Mesk{\'o}, Bertalan},
  title     = {The state of artificial intelligence-based {FDA}-approved
               medical devices and algorithms: an online database},
  journal   = {npj Digital Medicine},
  year      = {2020},
  volume    = {3},
  pages     = {118},
  doi       = {10.1038/s41746-020-00324-0},
}

@article{Veugelers1997,
  title={Internal R\&D expenditures and external technology sourcing},
  author={Veugelers, Reinhilde},
  year={1997},
  journal={Research Policy},
  doi={10.1016/s0048-7333(97)00019-x}
}

@article{Belderbos2004,
  author  = {Belderbos, Ren{\'e} and Carree, Martin and Lokshin, Boris},
  title   = {Cooperative {R\&D} and firm performance},
  year={2004},
  journal={Research Policy},
  doi={10.1016/j.respol.2004.07.003}
}

@article{Jaffe1989,
  title={Real Effects of Academic Research},
  author={Adam B. Jaffe},
  year={1989},
  journal={American Economic Review}
}

@article{AudretschFeldman1996,
  author  = {Audretsch, David B. and Feldman, Maryann P.},
  title   = {{R\&D} spillovers and the geography of innovation and production},
  journal = {The American Economic Review},
  year    = {1996}
}

@article{ZuckerDarby1998,
  author  = {Zucker, Lynne G. and Darby, Michael R. and Brewer, Marilynn B.},
  title   = {Intellectual human capital and the birth of {US} biotechnology
             enterprises},
  journal = {The American Economic Review},
  volume  = {88},
  number  = {1},
  pages   = {290--306},
  year    = {1998},
  doi     = {https://www.jstor.org/stable/116831}
}

@article{CockburnHenderson1998,
  title={Absorptive Capacity, Coauthoring Behavior, and the Organization of Research in Drug Discovery},
  author={Cockburn, Iain and Henderson, Rebecca M.},
  year={1998},
  journal={The Journal of Industrial Economics},
  doi={10.1111/1467-6451.00067}
}

@article{AroraGambardella1990,
  title={Complementarity and External Linkages: The Strategies of the Large Firms in Biotechnology},
  author={Ashish Arora and Alfonso Gambardella},
  year={1990},
  journal={Journal of Industrial Economics},
  doi={10.2307/2098345}
}

@article{Moncada2025,
  title={Digital adoption and human capital upscaling: a regional study of the manufacturing sector},
  author={Moncada, Roberto and Carbonero, Francesco and Geuna, Aldo and Riso, Luigi},
  year={2024},
  journal={Small Business Economics},
  doi={10.1007/s11187-024-00975-3}
}

@article{BodasFreitas2013,
  title={Finding the right partners: Institutional and personal modes of governance of university–industry interactions},
  author={Bodas Freitas, Isabel Maria and Geuna, Aldo and Rossi, Federica},
  year={2013},
  journal={Research Policy},
  doi={10.1016/j.respol.2012.06.007}
}

@article{Arundel2004,
  title={Proximity and the use of public science by innovative European firms},
  author={Arundel, Anthony and Geuna, Aldo},
  year={2004},
  journal={Economics of Innovation and New Technology},
  doi={10.1080/1043859092000234311}
}

@article{Laursen2006,
  author  = {Laursen, Keld and Salter, Ammon},
  title   = {Open for innovation: the role of openness in explaining
             innovation performance among {UK} manufacturing firms},
  journal = {Strategic Management Journal},
  volume  = {27},
  number  = {2},
  pages   = {131--150},
  year    = {2006}
}

@article{Hagedoorn2002,
  author  = {Hagedoorn, John},
  title   = {Inter-firm {R\&D} partnerships: an overview of major trends
             and patterns since 1960},
  journal = {Research Policy},
  volume  = {31},
  number  = {4},
  pages   = {477--492},
  year    = {2002}
}

@article{Miotti2003,
  author  = {Miotti, Luis and Sachwald, Fr{\'e}d{\'e}rique},
  title   = {Co-operative {R\&D}: why and with whom? {A}n integrated
             framework of analysis},
  journal = {Research Policy},
  volume  = {32},
  number  = {8},
  pages   = {1481--1499},
  year    = {2003}
}

@article{CassVeug2006,
  author  = {Cassiman, Bruno and Veugelers, Reinhilde},
  title   = {In Search of Complementarity in Innovation Strategy: Internal
             {R\&D} and External Knowledge Acquisition},
  journal = {Management Science},
  volume  = {52},
  number  = {1},
  pages   = {68--82},
  year    = {2006}
}

\clearpage
\appendix
\renewcommand{\thefigure}{S\arabic{figure}}
\renewcommand{\thetable}{S\arabic{table}}
\renewcommand{\theequation}{S\arabic{equation}}
\setcounter{figure}{0}
\setcounter{table}{0}
\setcounter{equation}{0}
\renewcommand{\thesection}{Appendix \Alph{section}}
\renewcommand{\theHsection}{Appendix.\Alph{section}}

This appendix provides additional materials supporting the main text. Section~\ref{app:robustness} reports robustness checks on the benchmark two-stage estimates. Section~\ref{app:desc} provides additional descriptive statistics and figures.

\section{Robustness checks and sensitivity analyses}
\label{app:robustness}

Table~\ref{tab:collab_decomp} decomposes the aggregate AI collaboration measure used throughout Section~\ref{sec:strategy} into four partner-type components: academic, company, healthcare, and other institutional affiliations, classified using the ROR typology described in Section~\ref{sec:data}. Column~(1) reproduces the all-firms 2SLS benchmark from Table~\ref{tab:benchmark_2stages} for reference. Columns~(2)--(5) re-estimate the same specification with the aggregate measure replaced by each category in turn.

Tables~\ref{tab:sens_I_2stages} and~\ref{tab:sens_II_2stages} present two sensitivity checks on the benchmark two-stage estimates. Sensitivity analysis~I drops the missing capital dummy from the specification, retaining only firms with non-missing capital stock. Sensitivity analysis~II removes all capital controls entirely. In both cases, the main result is preserved: collaborations in AI research retain a positive and significant effect on the probability of AI device introduction across all subsamples and both  estimators. The instrument set performs well throughout (Tables~\ref{tab:sens_I_tests} and~\ref{tab:sens_II_tests}), the only exception being that the null of the overidentification test is rejected for large firms and all firms in sensitivity analysis~I. Not taking into account the large number of missing values in the capital variable may thus weaken overidentification.

Table~\ref{tab:sens_III_ML_2step} compares ML and two-step estimation of the IV~Probit for large firms and the full sample. The two-step estimator yields qualitatively identical conclusions to the ML specification across both subsamples. Coefficients on collaborations in AI research are positive and significant under both approaches.

\begin{table}[htbp]
\centering
\caption{Collaboration in AI and AI innovation by type of partner (2SLS estimates, all firms)}
\label{tab:collab_decomp}
\setlength{\tabcolsep}{4pt}
\renewcommand{\arraystretch}{1.05}
\begin{adjustbox}{max width=\linewidth}
\footnotesize
\begin{tabular}{>{\raggedright\arraybackslash}p{5.5cm}ccccc}
\toprule
 & \multicolumn{5}{c}{\textbf{Type of collaboration partner}} \\
\cmidrule(lr){2-6}
 & \textbf{(1) Total (avg.)}
 & \textbf{(2) Academia}
 & \textbf{(3) Company}
 & \textbf{(4) Healthcare}
 & \textbf{(5) Others} \\
\midrule
\multicolumn{6}{l}{\textit{2\textsuperscript{nd} Stage --- Outcome: AI device dummy (introduction of $\geq$1 AI device)}} \\[2pt]
\rowcolor[gray]{0.95}
Collab.\ in AI research
  & $0.031^{***}$ & $0.053^{***}$ & $0.355^{**}$ & $0.155^{***}$ & $0.105^{***}$ \\
\rowcolor[gray]{0.95}
  & $(0.006)$ & $(0.020)$ & $(0.152)$ & $(0.056)$ & $(0.037)$ \\
Log $K$
  & $-0.003^{***}$ & $-0.003$ & $-0.003$ & $-0.003$ & $-0.003$ \\
  & $(0.001)$ & $(0.002)$ & $(0.002)$ & $(0.002)$ & $(0.002)$ \\
\rowcolor[gray]{0.95}
Missing $K$ dummy
  & $-0.279^{**}$ & $-0.290$ & $-0.311$ & $-0.307$ & $-0.278$ \\
\rowcolor[gray]{0.95}
  & $(0.109)$ & $(0.194)$ & $(0.198)$ & $(0.191)$ & $(0.190)$ \\
Log $L$
  & $0.003^{*}$ & $0.001$ & $0.001$ & $0.002$ & $0.001$ \\
  & $(0.002)$ & $(0.003)$ & $(0.003)$ & $(0.003)$ & $(0.003)$ \\
\rowcolor[gray]{0.95}
Log AI P.A.F.
  & $0.002^{*}$ & $0.009$ & $0.009$ & $0.009$ & $0.009$ \\
\rowcolor[gray]{0.95}
  & $(0.001)$ & $(0.006)$ & $(0.006)$ & $(0.006)$ & $(0.006)$ \\
Log Patenting
  & $0.001^{***}$ & $0.002$ & $0.004$ & $0.005$ & $0.005$ \\
  & $(0.0005)$ & $(0.028)$ & $(0.027)$ & $(0.027)$ & $(0.027)$ \\
\rowcolor[gray]{0.95}
Start-up
  & $0.016$ & $0.016$ & $0.013$ & $0.017$ & $0.017$ \\
\rowcolor[gray]{0.95}
  & $(0.011)$ & $(0.014)$ & $(0.015)$ & $(0.014)$ & $(0.014)$ \\
New entrant
  & $-0.004$ & $-0.002$ & $-0.003$ & $-0.002$ & $-0.004$ \\
  & $(0.010)$ & $(0.010)$ & $(0.010)$ & $(0.010)$ & $(0.010)$ \\
\rowcolor[gray]{0.95}
Country FE
  & Yes & Yes & Yes & Yes & Yes \\
Industry FE
  & Yes & Yes & Yes & Yes & Yes \\
\rowcolor[gray]{0.95}
Year FE
  & Yes & Yes & Yes & Yes & Yes \\[3pt]
\midrule
\multicolumn{6}{l}{\textit{1\textsuperscript{st} Stage --- Outcome: Average number of collaborations in AI research (per type of partner)}} \\[2pt]
\rowcolor[gray]{0.95}
Distance from nearest AI hub (km)
  & $-0.000022$ & $-0.000016$ & $-0.000005$ & $0.000003$ & $-0.000004$ \\
\rowcolor[gray]{0.95}
  & $(0.000029)$ & $(0.000014)$ & $(0.000003)$ & $(0.000006)$ & $(0.000009)$ \\
Log AI publications
  & $1.827^{***}$ & $1.046^{***}$ & $0.156^{***}$ & $0.362^{***}$ & $0.535^{***}$ \\
  & $(0.095)$ & $(0.104)$ & $(0.023)$ & $(0.035)$ & $(0.038)$ \\
Log $K$
  & $0.005$ & $0.001$ & $0.001$ & $0.002$ & $0.000$ \\
  & $(0.006)$ & $(0.006)$ & $(0.001)$ & $(0.002)$ & $(0.002)$ \\
\rowcolor[gray]{0.95}
Missing $K$ dummy
  & $0.439$ & $0.124$ & $0.083$ & $0.138$ & $-0.057$ \\
\rowcolor[gray]{0.95}
  & $(0.605)$ & $(0.576)$ & $(0.125)$ & $(0.241)$ & $(0.246)$ \\
Log $L$
  & $0.018^{**}$ & $0.009$ & $0.001$ & $-0.005^{*}$ & $0.000$ \\
  & $(0.008)$ & $(0.007)$ & $(0.002)$ & $(0.003)$ & $(0.003)$ \\
\rowcolor[gray]{0.95}
Log AI P.A.F.
  & $-0.003$ & $-0.007$ & $0.000$ & $-0.003$ & $-0.003$ \\
\rowcolor[gray]{0.95}
  & $(0.006)$ & $(0.014)$ & $(0.003)$ & $(0.007)$ & $(0.008)$ \\
Log Patenting
  & $0.008^{***}$ & $-0.020$ & $-0.009$ & $-0.022$ & $-0.031^{*}$ \\
  & $(0.002)$ & $(0.045)$ & $(0.008)$ & $(0.015)$ & $(0.017)$ \\
\rowcolor[gray]{0.95}
Start-up
  & $0.023$ & $0.006$ & $0.009$ & $-0.005$ & $-0.001$ \\
\rowcolor[gray]{0.95}
  & $(0.041)$ & $(0.027)$ & $(0.009)$ & $(0.009)$ & $(0.012)$ \\
New entrant
  & $-0.015$ & $-0.016$ & $0.002$ & $0.000$ & $0.012$ \\
  & $(0.025)$ & $(0.017)$ & $(0.006)$ & $(0.006)$ & $(0.008)$ \\
\rowcolor[gray]{0.95}
Country FE
  & Yes & Yes & Yes & Yes & Yes \\
Industry FE
  & Yes & Yes & Yes & Yes & Yes \\
\rowcolor[gray]{0.95}
Year FE
  & Yes & Yes & Yes & Yes & Yes \\[3pt]
\midrule
KP $F$-statistic (1st stage)
  & $186.8$ & $1{,}759.5$ & $543.8$ & $1{,}152.8$ & $1{,}429.4$ \\
\rowcolor[gray]{0.95}
Wu--Hausman ($p$-val.)
  & ${<}0.001$ & ${<}0.001$ & ${<}0.001$ & ${<}0.001$ & ${<}0.001$ \\
Observations
  & $8{,}774$ & $8{,}774$ & $8{,}774$ & $8{,}774$ & $8{,}774$ \\
\bottomrule
\end{tabular}
\end{adjustbox}
\begin{minipage}{\linewidth}
\smallskip\scriptsize
\textit{Notes:} $^{***}$ $p<0.01$, $^{**}$ $p<0.05$, $^{*}$ $p<0.1$.
Clustered robust standard errors in parentheses (clustered by firm).
FE rows for column~(1) report $p$-values of $F$-tests ($H_0$: FE not significant), as in Table~\ref{tab:benchmark_2stages};
``Yes'' indicates the set is included for columns~(2)--(5).
Column~(1) reproduces the all-firms 2SLS estimates from Table~\ref{tab:benchmark_2stages}. Columns~(2)--(5) replace the aggregate average collaboration count with the average number of distinct external co-author affiliations per year classified, respectively, as academic institutions (ROR types \textit{education} and \textit{research facility}), private-sector companies, healthcare organisations, and other affiliations (government bodies, non-profits, and unclassified entities). Each column is a separate 2SLS regression using the same instrument set ($\ln\mathrm{Pub}_{it}$ and $\mathrm{Distance}_{it}$) and controls. Kleibergen--Paap $F$-statistics all substantially exceed the Stock--Yogo critical value of 19.93, ruling out weak instruments throughout. Wu--Hausman tests confirm endogeneity of the collaboration measure at the 1\% level in all specifications.
\end{minipage}
\end{table}
\begin{table}[htbp]
\centering
\caption{Sensitivity analysis I (no ``missing capital'' dummy) — 2-stage estimates}
\label{tab:sens_I_2stages}
\setlength{\tabcolsep}{4pt}
\renewcommand{\arraystretch}{1.05}
\begin{adjustbox}{max width=\linewidth}
\footnotesize
\begin{tabular}{>{\raggedright\arraybackslash}p{5.2cm}cccccc}
\toprule
 & \multicolumn{2}{c}{\textbf{Small firms}}
 & \multicolumn{2}{c}{\textbf{Large firms}}
 & \multicolumn{2}{c}{\textbf{All firms}} \\
\cmidrule(lr){2-3}\cmidrule(lr){4-5}\cmidrule(lr){6-7}
 & \textbf{(1) 2SLS} & \textbf{(2) IV probit}
 & \textbf{(3) 2SLS} & \textbf{(4) IV probit}
 & \textbf{(5) 2SLS} & \textbf{(6) IV probit} \\
\midrule
\multicolumn{7}{l}{\textit{2\textsuperscript{nd} Stage — Outcome: AI dummy (introduction of $\geq$1 AI device)}} \\[2pt]
\rowcolor[gray]{0.95}
Collaborations in AI research (average)
  & $0.050^{***}$ & $0.251^{***}$
  & $0.023^{***}$ & $0.068$
  & $0.029^{***}$ & $0.155^{***}$ \\
\rowcolor[gray]{0.95}
  & $(0.009)$ & $(0.031)$ & $(0.009)$ & $(0.049)$ & $(0.006)$ & $(0.026)$ \\
Log $K$
  & $-0.0003$ & $0.010$
  & $-0.005^{**}$ & $-0.082^{***}$
  & $-0.002^{*}$ & $-0.019$ \\
  & $(0.001)$ & $(0.018)$ & $(0.002)$ & $(0.022)$ & $(0.001)$ & $(0.013)$ \\
\rowcolor[gray]{0.95}
Log $L$
  & $0.004$ & $0.065^{*}$
  & $0.003$ & $0.165^{*}$
  & $-0.0001$ & $0.062^{***}$ \\
\rowcolor[gray]{0.95}
  & $(0.003)$ & $(0.034)$ & $(0.006)$ & $(0.090)$ & $(0.002)$ & $(0.024)$ \\
Log AI P.A.F.
  & $0.006^{**}$ & $0.039^{***}$
  & $0.004^{***}$ & $0.075^{***}$
  & $0.003^{**}$ & $0.024^{**}$ \\
  & $(0.002)$ & $(0.014)$ & $(0.001)$ & $(0.022)$ & $(0.001)$ & $(0.010)$ \\
\rowcolor[gray]{0.95}
Log Patenting
  & $0.002^{***}$ & $0.033^{***}$
  & $-0.007^{***}$ & $-0.088^{***}$
  & $0.001$ & $0.007$ \\
\rowcolor[gray]{0.95}
  & $(0.0005)$ & $(0.008)$ & $(0.001)$ & $(0.018)$ & $(0.001)$ & $(0.007)$ \\
Start-up
  & $-0.006$ & $-0.038$
  & $\bar{\phantom{x}}$ & $\bar{\phantom{x}}$
  & $-0.015$ & $-0.044$ \\
  & $(0.013)$ & $(0.208)$ & & & $(0.013)$ & $(0.019)$ \\
Country FE ($p$-val.) & $0.090$ & $0.001$ & $0.000$ & $0.468$ & $0.158$ & $0.009$ \\
Industry FE ($p$-val.) & $0.000$ & $0.000$ & $0.000$ & $0.000$ & $0.000$ & $0.000$ \\
\rowcolor[gray]{0.95}
Year FE ($p$-val.) & $0.000$ & $0.000$ & $0.000$ & $0.000$ & $0.000$ & $0.000$ \\
\midrule
\multicolumn{7}{l}{\textit{1\textsuperscript{st} Stage — Outcome: Collaborations in AI research (avg.\ number)}} \\[2pt]
\rowcolor[gray]{0.95}
Log $K$
  & $0.001$ & $0.001$ & $0.014$ & $0.046^{**}$ & $0.002$ & $0.007$ \\
\rowcolor[gray]{0.95}
  & $(0.005)$ & $(0.005)$ & $(0.017)$ & $(0.019)$ & $(0.006)$ & $(0.005)$ \\
Log $L$
  & $0.037^{***}$ & $0.038^{***}$
  & $-0.010$ & $-0.118^{*}$
  & $0.024^{**}$ & $0.014$ \\
  & $(0.013)$ & $(0.013)$ & $(0.065)$ & $(0.061)$ & $(0.011)$ & $(0.009)$ \\
\rowcolor[gray]{0.95}
Log AI P.A.F.
  & $-0.002$ & $-0.002$ & $0.015$ & $-0.015$ & $-0.003$ & $-0.015^{**}$ \\
\rowcolor[gray]{0.95}
  & $(0.013)$ & $(0.013)$ & $(0.010)$ & $(0.011)$ & $(0.008)$ & $(0.007)$ \\
Log Patenting
  & $0.005^{**}$ & $0.005^{**}$
  & $0.019^{***}$ & $0.011$
  & $0.010^{***}$ & $0.006^{**}$ \\
  & $(0.003)$ & $(0.003)$ & $(0.007)$ & $(0.006)$ & $(0.003)$ & $(0.002)$ \\
\rowcolor[gray]{0.95}
Start-up
  & $0.069$ & $0.070$
  & $\bar{\phantom{x}}$ & $\bar{\phantom{x}}$
  & $0.026$ & $0.015$ \\
\rowcolor[gray]{0.95}
  & $(0.048)$ & $(0.048)$ & & & $(0.065)$ & $(0.052)$ \\
Distance from nearest AI hub (km)
  & $-0.00002$ & $-0.00001$
  & $-0.0001$ & $-0.00004$
  & $-0.00001$ & $-0.000001$ \\
  & $(0.00003)$ & $(0.00002)$ & $(0.0001)$ & $(0.00005)$ & $(0.00003)$ & $(0.00003)$ \\
\rowcolor[gray]{0.95}
Log AI publications
  & $3.900^{***}$ & $3.900^{***}$
  & $1.436^{***}$ & $2.691^{***}$
  & $1.860^{***}$ & $2.927^{***}$ \\
\rowcolor[gray]{0.95}
  & $(0.289)$ & $(0.288)$ & $(0.111)$ & $(0.176)$ & $(0.109)$ & $(0.271)$ \\
Country FE ($p$-val.) & $0.001$ & $0.001$ & $0.623$ & $0.700$ & $0.000$ & $0.000$ \\
Industry FE ($p$-val.) & $0.226$ & $0.219$ & $0.000$ & $0.238$ & $0.307$ & $0.255$ \\
\rowcolor[gray]{0.95}
Year FE ($p$-val.) & $0.075$ & $0.069$ & $0.116$ & $0.523$ & $0.004$ & $0.008$ \\
\midrule
$\text{atanh}\,\rho$
  & $\bar{\phantom{x}}$ & $-0.273^{***}$
  & $\bar{\phantom{x}}$ & $0.002$
  & $\bar{\phantom{x}}$ & $-0.126^{***}$ \\
  & & $(0.052)$ & & $(0.079)$ & & $(0.042)$ \\
\rowcolor[gray]{0.95}
$\text{Log}\,\sigma$
  & $\bar{\phantom{x}}$ & $0.029$
  & $\bar{\phantom{x}}$ & $0.184$
  & $\bar{\phantom{x}}$ & $0.102$ \\
\rowcolor[gray]{0.95}
  & & $(0.146)$ & & $(0.117)$ & & $(0.107)$ \\
\midrule
Observations & $3{,}709$ & $3{,}709$ & $1{,}591$ & $1{,}039$ & $5{,}300$ & $4{,}969$ \\
Firms        & $584$     & $584$     & $197$     & $136$     & $731$     & $705$ \\
\bottomrule
\end{tabular}
\end{adjustbox}
\begin{minipage}{\linewidth}
\smallskip\footnotesize
\textit{Notes:} Significance: $^{***}$ 1\%, $^{**}$ 5\%, $^{*}$ 10\%. Clustered robust standard errors in parentheses.
Country, Industry, and Year FE rows report $p$-values of $F$-tests (H$_0$: FE not significant).
The ``New entrant'' dummy is omitted in all sub-samples.
In 2SLS, the global $F$-test on all regressors is significant at 1\% in both stages. IV Probit models estimated by ML. Wald tests of exogeneity reject H$_0$: $\text{atanh}\,\rho=0$ at 1\%.
Marginal effects of ``avg.\ collaborations in AI'' at sample mean (IV Probit): $0.009^{***}$ $(0.002)$ — small firms; $0.004$ $(0.003)$ — large firms; $0.008^{***}$ $(0.001)$ — all firms.
\end{minipage}
\end{table}
\begin{table}[htbp]
\centering
\caption{Sensitivity analysis I: Post-estimation diagnostics for 2SLS models}
\label{tab:sens_I_tests}
\setlength{\tabcolsep}{12pt}
\renewcommand{\arraystretch}{1.1}
\begin{adjustbox}{max width=\linewidth}
\small
\begin{tabular}{lccc}
\toprule
 & \textbf{Small firms} & \textbf{Large firms} & \textbf{All firms} \\
 & \textbf{(1)} & \textbf{(2)} & \textbf{(3)} \\
\midrule
\multicolumn{4}{l}{\textbf{Underidentification Test} \textit{(H$_0$: Equation is underidentified)}} \\
~~Kleibergen-Paap LM statistic & $128.39$ & $241.58$ & $468.89$ \\
~~$p$-value                    & $0.000$  & $0.000$  & $0.000$  \\
\midrule
\multicolumn{4}{l}{\textbf{Weak Identification Test} \textit{(H$_0$: Instruments are weakly correlated)}} \\
~~Kleibergen-Paap $F$ statistic & $91.68$  & $84.57$  & $150.49$ \\
~~Critical value               & $19.93$  & $19.93$  & $19.93$  \\
\midrule
\multicolumn{4}{l}{\textbf{Overidentification Test} \textit{(H$_0$: Instruments are valid/exogenous)}} \\
~~Hansen $J$ statistic         & $3.17$   & $9.20$   & $11.03$  \\
~~$p$-value                    & $0.075$  & $0.002$  & $0.001$  \\
\bottomrule
\end{tabular}
\end{adjustbox}
\end{table}
\begin{table}[htbp]
\centering
\caption{Sensitivity analysis II (no control for capital) — 2-stage estimates}
\label{tab:sens_II_2stages}
\setlength{\tabcolsep}{4pt}
\renewcommand{\arraystretch}{1.05}
\begin{adjustbox}{max width=\linewidth}
\footnotesize
\begin{tabular}{>{\raggedright\arraybackslash}p{5.2cm}cccccc}
\toprule
 & \multicolumn{2}{c}{\textbf{Small firms}}
 & \multicolumn{2}{c}{\textbf{Large firms}}
 & \multicolumn{2}{c}{\textbf{All firms}} \\
\cmidrule(lr){2-3}\cmidrule(lr){4-5}\cmidrule(lr){6-7}
 & \textbf{(1) 2SLS} & \textbf{(2) IV probit}
 & \textbf{(3) 2SLS} & \textbf{(4) IV probit}
 & \textbf{(5) 2SLS} & \textbf{(6) IV probit} \\
\midrule
\multicolumn{7}{l}{\textit{2\textsuperscript{nd} Stage — Outcome: AI dummy (introduction of $\geq$1 AI device)}} \\[2pt]
\rowcolor[gray]{0.95}
Collaborations in AI research (average)
  & $0.046^{***}$ & $0.206^{***}$
  & $0.027^{***}$ & $0.111^{***}$
  & $0.030^{***}$ & $0.161^{***}$ \\
\rowcolor[gray]{0.95}
  & $(0.008)$ & $(0.029)$ & $(0.009)$ & $(0.042)$ & $(0.006)$ & $(0.023)$ \\
Log $L$
  & $0.009^{***}$ & $0.115^{***}$
  & $-0.006$ & $0.037$
  & $0.002$ & $0.083^{***}$ \\
  & $(0.002)$ & $(0.023)$ & $(0.004)$ & $(0.072)$ & $(0.001)$ & $(0.015)$ \\
\rowcolor[gray]{0.95}
Log AI P.A.F.
  & $0.004^{**}$ & $0.021^{**}$
  & $0.004^{***}$ & $0.063^{***}$
  & $0.002^{*}$ & $0.017^{**}$ \\
\rowcolor[gray]{0.95}
  & $(0.002)$ & $(0.010)$ & $(0.001)$ & $(0.021)$ & $(0.001)$ & $(0.008)$ \\
Log Patenting
  & $0.003^{***}$ & $0.034^{***}$
  & $-0.007^{***}$ & $-0.078^{***}$
  & $0.001^{***}$ & $0.018^{***}$ \\
  & $(0.0005)$ & $(0.006)$ & $(0.001)$ & $(0.017)$ & $(0.0005)$ & $(0.005)$ \\
\rowcolor[gray]{0.95}
Start-up
  & $0.024^{**}$ & $0.327^{***}$
  & $\bar{\phantom{x}}$ & $\bar{\phantom{x}}$
  & $0.016$ & $0.292^{***}$ \\
\rowcolor[gray]{0.95}
  & $(0.012)$ & $(0.119)$ & & & $(0.011)$ & $(0.112)$ \\
New entrant
  & $0.002$ & $0.038$
  & $-0.109$ & $-0.164$
  & $-0.005$ & $-0.007$ \\
  & $(0.009)$ & $(0.099)$ & $(0.078)$ & $(0.499)$ & $(0.011)$ & $(0.093)$ \\
Country FE ($p$-val.) & $0.213$ & $0.002$ & $0.003$ & $0.255$ & $0.596$ & $0.057$ \\
Industry FE ($p$-val.) & $0.000$ & $0.000$ & $0.000$ & $0.000$ & $0.000$ & $0.000$ \\
\rowcolor[gray]{0.95}
Year FE ($p$-val.) & $0.000$ & $0.000$ & $0.000$ & $0.000$ & $0.000$ & $0.000$ \\
\midrule
\multicolumn{7}{l}{\textit{1\textsuperscript{st} Stage — Outcome: Collaborations in AI research (avg.\ number)}} \\[2pt]
\rowcolor[gray]{0.95}
Log $L$
  & $0.036^{***}$ & $0.036^{***}$
  & $-0.005$ & $-0.029$
  & $0.024^{***}$ & $0.029^{***}$ \\
\rowcolor[gray]{0.95}
  & $(0.009)$ & $(0.009)$ & $(0.047)$ & $(0.051)$ & $(0.007)$ & $(0.006)$ \\
Log AI P.A.F.
  & $-0.007$ & $-0.007$ & $0.015$ & $-0.010$ & $-0.0001$ & $-0.009$ \\
  & $(0.007)$ & $(0.007)$ & $(0.010)$ & $(0.013)$ & $(0.006)$ & $(0.006)$ \\
\rowcolor[gray]{0.95}
Log Patenting
  & $0.004^{**}$ & $0.004^{**}$
  & $0.020^{***}$ & $0.013^{*}$
  & $0.008^{***}$ & $0.005^{***}$ \\
\rowcolor[gray]{0.95}
  & $(0.002)$ & $(0.002)$ & $(0.007)$ & $(0.007)$ & $(0.002)$ & $(0.002)$ \\
Start-up
  & $0.035$ & $0.035$
  & $\bar{\phantom{x}}$ & $\bar{\phantom{x}}$
  & $0.023$ & $0.025$ \\
  & $(0.037)$ & $(0.037)$ & & & $(0.041)$ & $(0.038)$ \\
\rowcolor[gray]{0.95}
New entrant
  & $-0.029$ & $-0.029$ & $0.581$ & $0.798$ & $-0.023$ & $-0.027$ \\
\rowcolor[gray]{0.95}
  & $(0.025)$ & $(0.025)$ & $(0.681)$ & $(0.902)$ & $(0.041)$ & $(0.025)$ \\
Distance from nearest AI hub (km)
  & $-0.00002$ & $-0.00002$
  & $-0.0006$ & $0.0001$
  & $-0.00002$ & $-0.00001$ \\
  & $(0.00002)$ & $(0.00002)$ & $(0.0006)$ & $(0.0001)$ & $(0.00002)$ & $(0.00002)$ \\
\rowcolor[gray]{0.95}
Log AI publications
  & $2.877^{***}$ & $2.878^{***}$
  & $1.446^{***}$ & $2.275^{***}$
  & $1.828^{***}$ & $2.596^{***}$ \\
\rowcolor[gray]{0.95}
  & $(0.270)$ & $(0.269)$ & $(0.101)$ & $(0.293)$ & $(0.095)$ & $(0.204)$ \\
Country FE ($p$-val.) & $0.000$ & $0.000$ & $0.093$ & $0.016$ & $0.000$ & $0.000$ \\
Industry FE ($p$-val.) & $0.262$ & $0.259$ & $0.000$ & $0.015$ & $0.462$ & $0.316$ \\
\rowcolor[gray]{0.95}
Year FE ($p$-val.) & $0.017$ & $0.009$ & $0.058$ & $0.488$ & $0.002$ & $0.004$ \\
\midrule
$\text{atanh}\,\rho$
  & $\bar{\phantom{x}}$ & $-0.144^{***}$
  & $\bar{\phantom{x}}$ & $-0.103$
  & $\bar{\phantom{x}}$ & $-0.108^{***}$ \\
  & & $(0.037)$ & & $(0.064)$ & & $(0.031)$ \\
\rowcolor[gray]{0.95}
$\text{Log}\,\sigma$
  & $\bar{\phantom{x}}$ & $-0.186$
  & $\bar{\phantom{x}}$ & $0.280^{**}$
  & $\bar{\phantom{x}}$ & $-0.087$ \\
\rowcolor[gray]{0.95}
  & & $(0.125)$ & & $(0.114)$ & & $(0.095)$ \\
\midrule
Observations & $7{,}131$ & $7{,}131$ & $1{,}643$ & $1{,}093$ & $8{,}774$ & $8{,}429$ \\
Firms        & $3{,}912$ & $3{,}912$ &   $234$ &   $166$ & $4{,}094$ & $4{,}063$ \\
\bottomrule
\end{tabular}
\end{adjustbox}
\begin{minipage}{\linewidth}
\smallskip\footnotesize
\textit{Notes:} Significance: $^{***}$ 1\%, $^{**}$ 5\%, $^{*}$ 10\%. Clustered robust standard errors in parentheses.

Country, Industry, and Year FE rows report $p$-values of $F$-tests (H$_0$: All parameters are 0, FE is not significant).
Since ``Start-up'' is always 0 among large firms, no coefficient is reported for that sub-sample.
In 2SLS, the global $F$-test is significant at 1\% in both stages. The IV Probit is estimated by ML. Wald tests of exogeneity reject H$_0$: $\text{atanh}\,\rho=0$ at 1\% among small firms and in all firms (full sample). Marginal effects of ``avg.\ collaborations in AI'' at sample mean (IV Probit): $0.012^{***}$ $(0.002)$ — small firms; $0.008^{**}$ $(0.003)$ — large firms; $0.011^{***}$ $(0.002)$ — all firms.
\end{minipage}
\end{table}
\begin{table}[htbp]
\centering
\caption{Sensitivity analysis II: Post-estimation diagnostics for 2SLS models}
\label{tab:sens_II_tests}
\setlength{\tabcolsep}{12pt}
\renewcommand{\arraystretch}{1.1}
\begin{adjustbox}{max width=\linewidth}
\small
\begin{tabular}{lccc}
\toprule
 & \textbf{Small firms} & \textbf{Large firms} & \textbf{All firms} \\
 & \textbf{(1)} & \textbf{(2)} & \textbf{(3)} \\
\midrule
\multicolumn{4}{l}{\textbf{Underidentification Test} \textit{(H$_0$: Equation is underidentified)}} \\
~~Kleibergen-Paap LM statistic & $152.53$ & $262.27$ & $463.57$ \\
~~$p$-value                    & $0.000$  & $0.000$  & $0.000$  \\
\midrule
\multicolumn{4}{l}{\textbf{Weak Identification Test} \textit{(H$_0$: Instruments are weakly correlated)}} \\
~~Kleibergen-Paap $F$ statistic & $58.95$  & $103.22$ & $187.92$ \\
~~Critical value               & $19.93$  & $19.93$  & $19.93$  \\
\midrule
\multicolumn{4}{l}{\textbf{Overidentification Test} \textit{(H$_0$: Instruments are valid/exogenous)}} \\
~~Hansen $J$ statistic         & $1.654$  & $0.775$  & $4.962$  \\
~~$p$-value                    & $0.198$  & $0.379$  & $0.026$  \\
\bottomrule
\end{tabular}
\end{adjustbox}
\end{table}
\begin{table}[htbp]
\centering
\caption{Sensitivity analysis III (ML vs.\ 2-step estimation of IV probit)}
\label{tab:sens_III_ML_2step}
\setlength{\tabcolsep}{5pt}
\renewcommand{\arraystretch}{1.05}
\begin{adjustbox}{max width=\linewidth}
\footnotesize
\begin{tabular}{lcccccc}
\toprule
 & \multicolumn{3}{c}{\textbf{Large firms}} & \multicolumn{3}{c}{\textbf{All firms}} \\
\cmidrule(lr){2-4} \cmidrule(lr){5-7}
 & \textbf{(1)} & \textbf{(2)} & \textbf{(3)} & \textbf{(4)} & \textbf{(5)} & \textbf{(6)} \\
 & \textbf{2SLS} & \textbf{IV probit (ML)} & \textbf{IV probit (2-step)} & \textbf{2SLS} & \textbf{IV probit (ML)} & \textbf{IV probit (2-step)} \\
\midrule
\multicolumn{7}{l}{\textit{2\textsuperscript{nd} Stage Equation — Outcome variable: AI dummy (Introduction of $\ge 1$ AI device)}} \\[2pt]
\midrule
\rowcolor[gray]{0.95}
Collaborations in AI      & $0.028^{***}$ & $0.109^{**}$  & $0.197^{***}$  & $0.031^{***}$ & $0.165^{***}$  & $0.150^{***}$  \\
\rowcolor[gray]{0.95}
~~research (average)       & $(0.009)$     & $(0.045)$     & $(0.053)$      & $(0.006)$     & $(0.023)$      & $(0.024)$      \\
Log $K$                    & $-0.005^{**}$ & $-0.087^{***}$ & $-0.057^{**}$  & $-0.003^{***}$ & $-0.026^{**}$  & $-0.026^{**}$  \\
                           & $(0.002)$     & $(0.023)$     & $(0.021)$      & $(0.001)$     & $(0.012)$      & $(0.011)$      \\
\rowcolor[gray]{0.95}
$K$ dummy                  & $-0.449^{*}$  & $-8.094^{***}$ & $-4.898^{**}$  & $-0.279^{**}$ & $-2.931^{**}$  & $-2.966^{***}$ \\
\rowcolor[gray]{0.95}
                           & $(0.260)$     & $(2.422)$     & $(2.352)$      & $(0.109)$     & $(1.227)$      & $(1.134)$      \\
Log $L$                    & $0.001$       & $0.182^{**}$  & $-0.009$       & $0.003^{*}$   & $0.085^{***}$  & $0.043^{**}$   \\
                           & $(0.005)$     & $(0.087)$     & $(0.062)$      & $(0.002)$     & $(0.018)$      & $(0.018)$      \\
\rowcolor[gray]{0.95}
Log AI P.A.F.              & $0.005^{**}$  & $0.068^{***}$  & $0.070^{***}$  & $0.002^{*}$   & $0.018^{**}$   & $0.016^{**}$   \\
\rowcolor[gray]{0.95}
                           & $(0.001)$     & $(0.021)$     & $(0.022)$      & $(0.001)$     & $(0.008)$      & $(0.008)$      \\
Log Patenting              & $-0.007^{***}$ & $-0.086^{***}$ & $-0.095^{***}$ & $0.001^{***}$ & $0.017^{***}$  & $0.018^{***}$  \\
                           & $(0.001)$     & $(0.017)$     & $(0.018)$      & $(0.0005)$    & $(0.005)$      & $(0.005)$      \\
\rowcolor[gray]{0.95}
Start-up                   & --            & --            & --            & $0.016$       & $0.301^{***}$  & $0.255^{**}$   \\
\rowcolor[gray]{0.95}
                           &               &               &               & $(0.011)$     & $(0.112)$      & $(0.117)$      \\
New entrant                & $-0.173^{*}$  & $-0.235^{*}$  & $-1.489$      & $-0.004$      & $0.024$        & $0.016$        \\
                           & $(0.090)$     & $(0.690)$     & $(0.863)$      & $(0.010)$     & $(0.094)$      & $(0.101)$      \\
\rowcolor[gray]{0.95}
Country FE ($p$-val.)      & $0.001$       & $0.535$       & $0.717$        & $0.442$       & $0.021$        & $0.123$        \\
Industry FE ($p$-val.)     & $0.000$       & $0.000$       & $0.000$        & $0.000$       & $0.000$        & $0.000$        \\
Year FE ($p$-val.)         & $0.000$       & $0.000$       & $0.000$        & $0.000$       & $0.000$        & $0.000$        \\
\midrule
\multicolumn{7}{l}{\textit{1\textsuperscript{st} Stage Equation — Outcome variable: Collaborations in AI research (average number)}} \\[2pt]
\midrule
\rowcolor[gray]{0.95}
Log $K$                    & $0.017$       & $0.044^{**}$  & $0.014$        & $0.005$       & $0.009^{*}$    & $0.005$        \\
\rowcolor[gray]{0.95}
                           & $(0.016)$     & $(0.019)$     & $(0.018)$      & $(0.006)$     & $(0.005)$      & $(0.005)$      \\
$K$ dummy                  & $1.436$       & $4.252^{**}$  & $0.968$        & $0.439$       & $0.825$        & $0.439$        \\
                           & $(1.723)$     & $(1.972)$     & $(2.041)$      & $(0.605)$     & $(0.527)$      & $(0.539)$      \\
\rowcolor[gray]{0.95}
Log $L$                    & $-0.022$      & $-0.097$      & $-0.006$       & $0.018^{**}$  & $0.019^{***}$  & $0.018^{**}$   \\
\rowcolor[gray]{0.95}
                           & $(0.056)$     & $(0.069)$     & $(0.050)$      & $(0.008)$     & $(0.007)$      & $(0.009)$      \\
Log AI P.A.F.              & $0.013$       & $-0.011$      & $0.013$        & $-0.003$      & $-0.009$       & $-0.0003$      \\
                           & $(0.011)$     & $(0.013)$     & $(0.014)$      & $(0.006)$     & $(0.006)$      & $(0.004)$      \\
\rowcolor[gray]{0.95}
Log Patenting              & $0.021^{***}$ & $0.015^{**}$  & $0.019$        & $0.008^{***}$ & $0.005^{***}$  & $0.008^{***}$  \\
\rowcolor[gray]{0.95}
                           & $(0.007)$     & $(0.007)$     & $(0.012)$      & $(0.002)$     & $(0.002)$      & $(0.002)$      \\
Start-up                   & --            & --            & --            & $0.023$       & $0.024$        & $0.023$        \\
                           &               &               &               & $(0.041)$     & $(0.038)$      & $(0.060)$      \\
\rowcolor[gray]{0.95}
New entrant                & $0.954$       & $1.190$       & $1.191$        & $-0.015$      & $-0.015$       & $-0.015$       \\
\rowcolor[gray]{0.95}
                           & $(0.720)$     & $(0.964)$     & $(0.763)$      & $(0.025)$     & $(0.025)$      & $(0.052)$      \\
Distance to AI hub (km)    & $-0.0001$     & $0.0001$      & $-0.0001$      & $-0.00002$    & $-0.00001$     & $-0.00002$     \\
                           & $(0.0001)$    & $(0.0001)$    & $(0.0001)$     & $(0.00002)$   & $(0.00002)$    & $(0.00003)$    \\
\rowcolor[gray]{0.95}
Log AI publications        & $1.438^{***}$ & $2.282^{***}$ & $1.407^{***}$  & $1.827^{***}$ & $2.596^{***}$  & $1.827^{***}$  \\
\rowcolor[gray]{0.95}
                           & $(0.100)$     & $(0.296)$     & $(0.069)$      & $(0.095)$     & $(0.203)$      & $(0.031)$      \\
Country FE ($p$-val.)      & $0.560$       & $0.539$       & $0.129$        & $0.000$       & $0.010$        & $0.000$        \\
Industry FE ($p$-val.)     & $0.000$       & $0.075$       & $0.000$        & $0.418$       & $0.303$        & $0.452$        \\
Year FE ($p$-val.)         & $0.057$       & $0.457$       & $0.122$        & $0.002$       & $0.002$        & $0.002$        \\
\midrule
\multicolumn{7}{l}{\textit{Ancillary Model Parameters}} \\[2pt]
\midrule
\rowcolor[gray]{0.95}
$\operatorname{atanh}\rho$  & --            & $-0.065$      & --            & --            & $-0.112^{***}$ & --            \\
\rowcolor[gray]{0.95}
                           &               & $(0.068)$     &               &               & $(0.031)$      &               \\
$\log\sigma$               & --            & $0.276^{**}$  & --            & --            & $-0.088$       & --            \\
                           &               & $(0.113)$     &               &               & $(0.095)$      &               \\
\midrule
Observations               & $1{,}643$     & $1{,}093$     & $1{,}502$      & $8{,}774$     & $8{,}429$      & $8{,}774$      \\
Firms                      & $234$         & $166$         & $208$          & $4{,}094$     & $4{,}063$      & $4{,}094$      \\
\bottomrule
\end{tabular}
\end{adjustbox}
\begin{minipage}{\linewidth}
\smallskip \tiny
\textit{Notes:} Significance levels: $^{***}$ 1\%, $^{**}$ 5\%, $^{*}$ 10\%. Clustered robust standard errors in parentheses for 2SLS and ML estimates of the IV Probit; unclustered asymptotic standard errors in parentheses for the 2-step IV Probit. 

The FE rows display $p$-values of $F$-tests ($H_0$: FE coefficients are jointly zero). The ``Start-up'' dummy variable is always 0 within the large firms sub-sample and is therefore omitted from the regressors in this sub-sample. With 2SLS, goodness-of-fit $F$-tests systematically reject the null at the 1\% significance level in both stages. The IV Probit are estimated by Maximum Likelihood (ML) in Columns (2) and (5), and with Newey's Minimum $\chi^2$ two-step procedure in Columns (3) and (6). Parameters $\rho$ and $\sigma$ are the correlation coefficient between the first- and second-stage errors of the IV Probit and the standard deviation of the first-stage error, respectively. Wald tests on $\rho$ reject the null hypothesis of exogeneity of the instrumented regressor at the 1\% level in the full sample ("all firms").
\end{minipage}
\end{table}

\clearpage

\section{Additional descriptive statistics}
\label{app:desc}

\begin{figure}[httb]
  \centering
  \includegraphics[width=\textwidth]{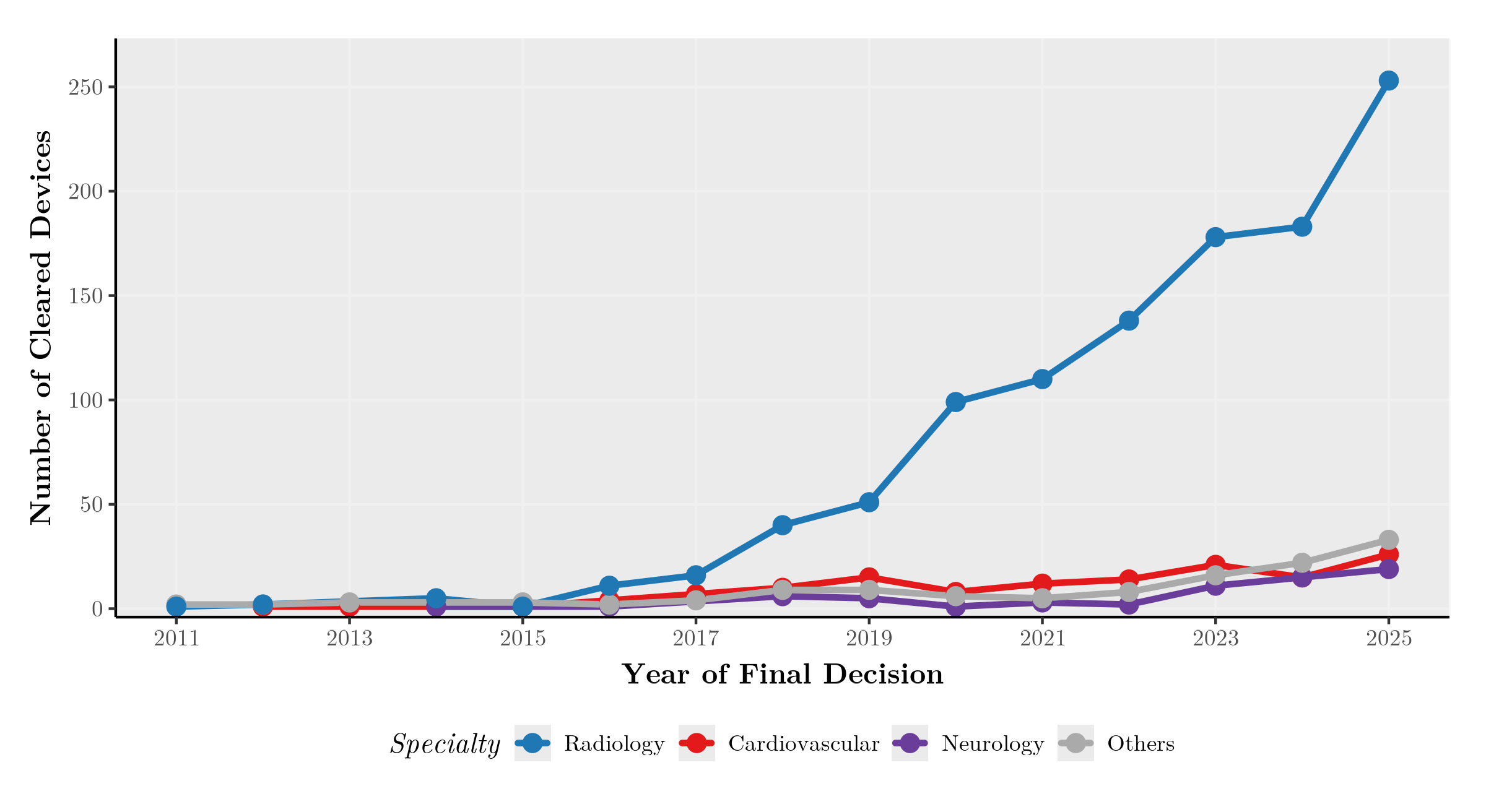}
  \caption{\textbf{AI-Enabled Medical Device Clearance by Specialty.}}
  \label{fig:ai_device_specialty}
\end{figure}

\end{document}